\documentclass{aa}

\usepackage{graphicx}
\usepackage{tabularx}
\usepackage{xcolor}

\usepackage[varg]{txfonts}

\newcommand{\gc}{\bf{}}
\renewcommand{\gc}{} 

\begin{document}

   \title{A cavity and further radial substructures in the disk around HD~97048}

     \author{G. van der Plas \inst{1,2}
          \and C. M. Wright \inst{3}
          \and F. M\'enard \inst{4,5}
          \and S. Casassus \inst{1,2}
          \and H. Canovas \inst{6}
          \and C. Pinte \inst{4,5}
          \and S. T. Maddison \inst{7}
          \and K. Maaskant \inst{8}
          \and H. Avenhaus \inst{1,2}
          \and L. Cieza \inst{2,9}
          \and S. Perez \inst{1,2}
          \and C. Ubach \inst{10} }

    \institute{Departamento de Astronomia, Universidad de Chile, Casilla 36-D, Santiago, Chile
        \and Millenium Nucleus Protoplanetary Disks in ALMA Early Science, Universidad de Chile, Casilla 36-D, Santiago, Chile
        \and School of Physical, Environmental and Mathematical Sciences, UNSW@ADFA, Canberra, ACT 2600, Australia
        \and CNRS, IPAG, F-38000 Grenoble, France 
        \and Univ. Grenoble Alpes, IPAG, F-38000 Grenoble, France 
        \and Departamento de F\'isica Te\'orica, Universidad Aut\'onoma de Madrid, Cantoblanco 28049 Madrid, Spain.
        \and Centre for Astrophysics \& Supercomputing, Swinburne University, PO Box 218, Hawthorn, VIC 3122, Australia
        \and ASML, De Run 6501, 5504 DR Veldhoven, Netherlands.
        \and Nucleo de Astronomia, Facultad de Ingenieria, Universidad Diego Portales, Av Ejercito 441, Santiago, Chile
        \and NRAO, 520 Edgemont Rd, Charlottesville, VA 22903-2475, USA}

   \date{Received tbd; accepted tbd}

    \abstract
   {Gaps, cavities and rings in circumstellar disks are signposts of disk evolution and planet-disk interactions. We follow the recent suggestion that Herbig Ae/Be disks with a flared disk harbour a cavity, and investigate the disk around HD~97048.}
   {We aim to resolve the 34$\pm$ 4 au central cavity predicted by Maaskant et al. (2013) and to investigate the structure of the disk.}
   {We image the disk around HD~97048 using ALMA at 0.85~mm and 2.94~mm, and ATCA (multiple frequencies) observations. Our observations also include the \element[ ][12]{CO} J=1-0, \element[ ][12]{CO} J=3-2 and \element[+][]{HCO} J=4-3 emission lines.}
   {A central cavity in the disk around HD~97048 is resolved with a 40-46 au radius. Additional radial structure present in the surface brightness profile  can be accounted for either by an opacity gap at ~90 au or by an extra emitting ring at ~150 au. The continuum emission tracing the dust in the disk is detected out to 355 au. The \element[ ][12]{CO} J=3-2 disk is detected 2.4 times farther out. The \element[ ][12]{CO} emission can be traced down to $\approx$ 10 au scales. Non-Keplerian kinematics are detected inside the cavity via the \element[+][]{HCO} J=4-3 velocity map. The mm spectral index measured from ATCA observations suggests that grain growth has occurred in the HD~97048 disk. Finally, we resolve a highly inclined disk out to 150 au around the nearby 0.5~$M_{\sun}$ binary ISO-ChaI 126.}
   {The data presented here reveal a cavity in the disk of HD 97048, and prominent radial structure in the surface brightness. The cavity size varies for different continuum frequencies and gas tracers. The gas inside the cavity follows non-Keplerian kinematics seen in \element[+][]{HCO} emission. The variable cavity size along with the kinematical signature suggests the presence of a substellar companion or massive planet inside the cavity.}

\maketitle
   
\section{Introduction}

Protoplanetary disks are the birth environments of planetary systems. How these planets form in their disks is an ongoing topic of debate, which is informed by an increasing number of disks that show various degrees of dispersal such as opacity cavities (transitional disks) and opacity gaps (pre-transitional disks) \citep[e.g.][]{2011ARA&A..49...67W}. Examples of such disks with directly imaged cavities at (sub) mm wavelengths include  HD~100546 \citep{2014ApJ...791L...6W}, Sz~91 \citep{2015ApJ...805...21C, 2016MNRAS.458L..29C}, LkCa~15 \citep{2006A&A...460L..43P, 2011ApJ...742L...5A}, HD~142527 \citep{2013Natur.493..191C} and SAO~206462 \citep{2009ApJ...704..496B}. 
The common denominator between these disks is that their structure can be described by one large cavity or a broad ring of dust grains at reasonably large radii and with large ring widths (at least tens of au in radii and width) or with a pile-up of large dust in narrow rings. The gaps and/or cavities in these disks are not empty: they contain both smaller dust grains, as traced by scattered light imaging \citep[e.g. ][]{2012ApJ...745....5K, 2014ApJ...781...87A}, and gas, as traced by rotational \citep{2015ApJ...798...85P, 2015A&A...579A.106V} and ro-vibrational carbon monoxide (CO) lines \citep{2009A&A...500.1137V, 2011ApJ...733...84P}. Recently, long baseline Atacama Large Millimeter Array (ALMA) observations of HL Tau \citep{2015ApJ...808L...3A} and TW Hya \citep{2016ApJ...820L..40A, 2016ApJ...819L...7N, 2016arXiv160500289T}  have demonstrated that these disks show a rich substructure of many concentric rings and gaps at scales as small as 1 au when observed at very high spatial resolution. Indeed, it is possible that most disks contain similar detailed structures which have not yet observed \citep{2016ApJ...818L..16Z}.

Disks around Herbig Ae/Be (HAeBe) stars have historically been split into group I and group II. Group I sources have been interpreted as hosting gas-rich protoplanetary disks with a flared, bright dust surface; whereas the dust in group II disks is assumed to have settled towards the mid-plane and are therefore weak in mid- to far-infrared emission \citep{2001A&A...365..476M, 2004A&A...417..159D}. Recent modeling of resolved observations of group I sources suggest that their bright infrared emission can be attributed to the large vertical walls that exists as a consequence of large dust cavities \citep{2012ApJ...752..143H,2013A&A...555A..64M}.

HD~97048 is a 3 Myr \citep{2006Sci...314..621L}, 2.5 M$_\sun$ \citep{2007A&A...470..625D} HAeBe  with spectral type A0 at a distance of 158$^{+17}_{-14}$ parsec \citep{2007A&A...474..653V}. Its Spectral Energy Distribution (SED) is classified as group I, is bright in the mid- to far-IR, and rich in PAH features, but lacks any sign of amorphous and crystalline silicate features \citep{2014A&A...563A..78M}. The mass accretion rate onto the star is low with an upper limit of log(M$_{acc}$) $\leq$ -8.16 M$_{\odot}$ yr$^{-1}$ \citep{2015MNRAS.453..976F}. 

The disk around HD~97048 is exceptionally bright and is one of only two HAeBe disks in which near-IR 1-0 S(1)  \citep{2011A&A...533A..39C} and Mid-IR \citep{2009ApJ...695.1302M} H$_2$ emission has been detected. The outer disk has been resolved in the PAH bands and shows a typical flaring geometry, with a flaring index of 1.26$\pm$0.05 \citep{2006Sci...314..621L}, and an inclined disk geometry with the eastern side farther from us. The disk has been very well studied using CO emission, and shows an 11 au cavity in ro-vibrational emission \citep{2009A&A...500.1137V}, but no detection in overtone emission \citep[][]{2015A&A...574A..75V}. The rotational CO ladder is richly populated as detected by {\textit Herschel} \citep[][]{2013A&A...559A..84M, 2014MNRAS.444.3911V, 2016arXiv160402055F}. In the (sub) mm bands, this disk has only been detected using single dish observations and has never resolved \citep{1998A&A...336..565H,2011PhDT.......219P, 2014AJ....148...47H}. The disk has, however, been resolved by polarimetric differential imaging of polarized scattered light \citep{2012A&A...538A..92Q}, showing a bright disk surface between $\approx$ 0.1\arcsec–1.0\arcsec ($\approx$16–160 AU), but no evidence for a disk cavity. \citet{2013A&A...555A..64M} also resolve the disk in the Q band (20 $\mu$m) spectrum and find that a large gap should be present in the disk between 2 and 34$^{+4}_{-4}$ au.

In this manuscript we present resolved (sub) mm observations of the disk around HD~97048 obtained with ALMA and the Australia Telescope Compact Array (ATCA)\footnote{The Australia Telescope Compact Array is part of the Australia Telescope which is funded by the Commonwealth of Australia for operation as a National Facility managed by CSIRO.}. We describe the observations and data reduction in \S \ref{sec:obs+data}, and present the results in \S \ref{sec:results}. We discuss these results and put them into context with previous results in \S \ref{sec:discussion}, and present our conclusions in  \S \ref{sec:conclusion}.

\begin{figure*}
   \centering
   \includegraphics[width=\hsize]{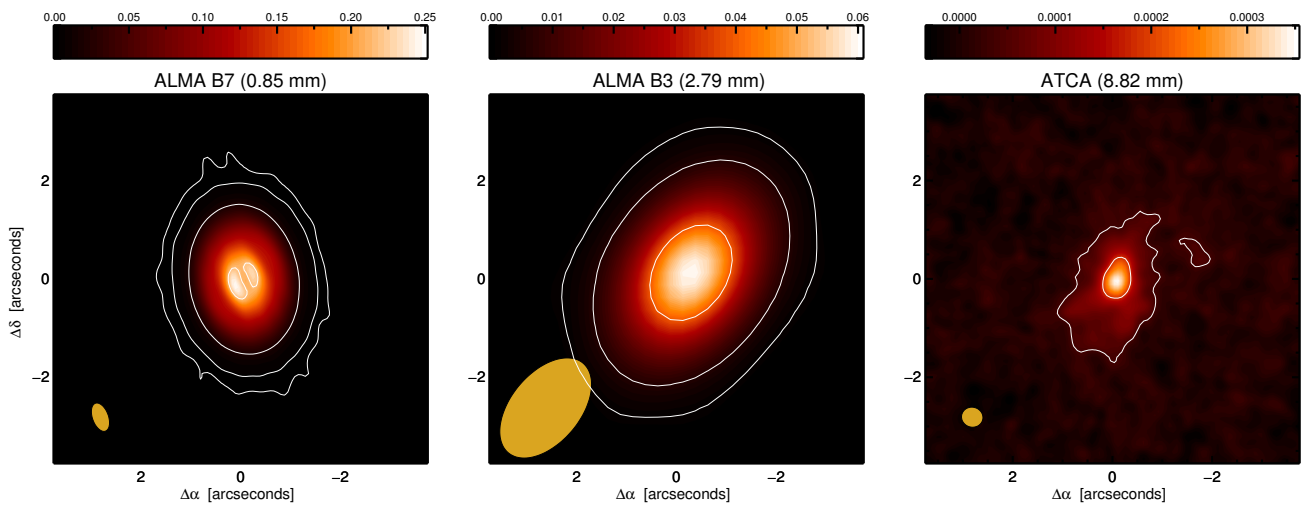}
     \caption{Images of \object{HD 97048} for the ALMA band 7 (left panel) and band 3 (central panel) and the combined ATCA 33+35 GHz (right panel) observations, reconstructed using uniform (ALMA) and natural (ATCA) weighting. The intensity scale for all images is in units of Jy/beam. Over plotted are contours with [3, 15, 100 and 1100] times the RMS value of respectively 0.20, 0.18 and 0.010 mJy beam$^{-1}$. The beam is shown in orange in the bottom left of each panel.}
    \label{fig:continuum_all}
\end{figure*}

\section{Observations and data reduction}\label{sec:obs+data}

\begin{table*}
\begin{center}
\scriptsize

\caption{Observational details for the ALMA and ATCA observations. Note that as a quality indicator for the observational conditions, we give the pwv value for the ALMA data and the RMS path length error -- measured at 21~GHz along a 200~m path -- for the ATCA observations. The latter has a decreasing influence on the data quality (i.e. phase stability) as both the frequency and baseline length decreases. \label{table_obs}} \smallskip
\begin{minipage}[t]{\textwidth}
\noindent\begin{tabularx}{\columnwidth}{@{\extracolsep{\stretch{1}}}*{9}{c}@{}}
\hline\hline
ID & UT Date & No. Antennas & Time on Target & Baseline Range & pwv &  \multicolumn{3}{c}{Calibrators} \\ 
 &  & (array) &  & (m) & ($\mu$m) & Flux & Bandpass & Gain  \\
\hline
ALMA Band 3 & 2013 Dec 1	& 25 & 1m12.6s & 15.8 to 462.9 &  535 & Ganymede & J0137-2934 & J1058-8003\\
ALMA Band 7 & 2015 May 22	& 36 & 7m10s & 21.6 to 555.5 & 687 & J1107-448 & J1337-1257 & J1058-8003 \\
\hline
   &  &   &    &  & RMS  & & & \\
\hline

ATCA 95/97 GHz & 2004 Oct 14 & 5 (H214C) & 30m  &  82.4--246.8  & N/A & Uranus &  B0537-441  &  B1057-797 \\
ATCA 93/95 GHz & 2009 Jul 30 & 5 (1.5A) & 1h30m & 153.1--1469.4 & 80--180 & Uranus & B1253-055 & B1057-797 \\
ATCA 43/45 GHz & 2009 Aug 02 & 6 (1.5A) & 3h    & 153.1--4469.4  & 50--100 & Uranus & B1921-293 & B1057-797 \\
ATCA 43/45 GHz & 2009 Aug 03 & 6 (1.5A)  & 70m   & 153.1--4469.4 & 50 & Uranus & B1921-293 & B1057-797 \\
ATCA 43/45 GHz & 2009 Dec 08 & 6 (EW352) & 2h38m & 30.6--4438.8 & 300--600 & B1057-797 & B0537-441 & B1057-797 \\
ATCA 33/35 GHz & 2010 May 01 & 6 (6A) & 3h20m & 336.7--5938.8 & 150--500  & B1934-638 & B1253-055 & B1057-797 \\
ATCA 33/35 GHz & 2010 Jun 10 & 6 (6C) & 45m  & 153.1--6000.0 & 150 & B1934-638 & B1253-055 & B1057-797 \\
ATCA 43/45 GHz & 2010 Jul 18 & 6 (EW352) & 1h17m & 30.6--4438.8 & 300--640 & Uranus & B0537-441 & B1057-797 \\
ATCA 93/95 GHz & 2010 Jul 19 & 5 (EW352) & 1h  & 30.6--352.0 & 70--240 & Uranus & B0537-441 & B1057-797 \\
ATCA 43/45 GHz & 2010 Jul 21 & 6 (EW352) & 1h20m & 30.6--4438.8 & 110--230 & Uranus & B2223-052 & B1057-797 \\
ATCA 93/95 GHz & 2010 Jul 21 & 5 (EW352) & 45m & 30.6--352.0 & 90--130 & Uranus & B2223-052 & B1057-797 \\
ATCA 91/97 GHz & 2010 Oct 06 & 5 (H214) & 2h47m & 82.4--246.8 & 140--290 & Uranus & B1921-293 & B1057-797 \\
ATCA 18/24 GHz & 2010 Oct 08 & 6 (H214) & 20m  & 82.4--4500.0 & 140--160 & J1047-6217 & B0537-441 & B1057-797 \\
ATCA 5.5/9 GHz & 2011 May 30 & 6 (H214) & 1h  & 82.4--4500.0 & 90--200 & B1934-638 & B0537-441 & B1057-797 \\
ATCA 5.5/9 GHz & 2011 May 31 & 6 (H214) & 30m  & 82.4--4500.0 & 90--150 & B1934-638 & B0537-441 & B1057-797 \\
ATCA 33/35 GHz & 2011 May 31 & 6 (H214) & 40m  & 82.4--4500.0 & 130--200 & B1934-638 & B0537-441 & B1057-797 \\
ATCA 17/19 GHz & 2011 Jun 01 & 6 (H214) & 20m  & 82.4--4500.0 & 170--250 & B1934-638 & B1147-6753 & B1057-797 \\
ATCA 17/19 GHz & 2011 Jul 09 & 6 (H214) & 30m  & 82.4--4500.0 & 100--200 & B1057-797 & B1921-293 & B1057-797 \\
ATCA 5.5/9 GHz & 2012 Jul 14 & 6 (H168) & 20m  & 61.2--4469.4 & 600--800 & B1057-797 & B0537-441 & B1057-797 \\
ATCA 33/35 GHz & 2013 Jul 27 & 6 (6A) & 3h20m & 336.7 to 5938.8 & 100--800 & B1934-638 & B0420-014 & B1057-797 \\
ATCA 33/35 GHz & 2014 Aug 30 & 6 (6B) & 9.46h  & 214.3 to 5969.4 & 100--300 & B1934-638 & B0537-441 & B1057-797 \\
ATCA 38/40 GHz & 2014 Aug 31 & 6 (6B) & 9.32h & 214.3 to 5969.4 & 50--600 & B1934-638 & B0537-441 & B1057-797 \\

\hline

\end{tabularx}
\end{minipage}
\end{center}
\end{table*}


\subsection{ATCA data}

High spatial resolution 7--9~mm imaging observations of HD~97048 were conducted on 30 and 31 August 2014 under project C1794. Full synthesis tracks were performed in the 6B east-west configuration, with baselines ranging between 214.3~m to 5969.4~m. During the observations the weather conditions were good on 30 August, with the RMS path length error (or seeing) $\leq$ 300~$\mu$m, and fair during 31 August, with the seeing generally below 300~$\mu$m for much of the track but spiking to 600~$\mu$m for several hours in the middle of the track.   For the 30 August 2014 track, the two sidebands of the 7~mm receiver were centred on 33 and 35~GHz, and on 38 and 40~GHz on 31 August.

In addition to the above imaging-oriented observations, multiple ATCA 3~mm, 7~mm, 9~mm, 1.7~cm, 3.3~cm and 5.45~cm observations were conducted between 14 October 2004 and 28 July 2013. All these were either only partial tracks in an E-W configuration -- sometimes as fillers -- under projects C996, C1173 and C2094, or relatively short observations in a compact hybrid configuration under the same project codes, plus C2534 and C2426. They provide reasonable flux estimates but low fidelity images due to their poor uv coverage. This itself is a product of shorter integration times (and hence extended beams in the E-W arrays), and that HD~97048 was observed on several occasions at elevations down to $\leq$ 20\degr, where the efficiency is poor. 

For the 3~mm observations of October 2004, the old ATCA correlator was used and set to continuum mode, each bandwidth being 128~MHz wide with 32 channels. From August 2009 onwards, all the observations used the Compact Array Broadband Backend, or CABB \citep{2011MNRAS.416..832W}, which provides a bandwidth of 2~GHz with 2048 $\times$ 1~MHz channels per sideband.  

The bandpass and flux calibrators were observed for $\sim15$~mins each, and the phase calibrator was observed every 5--10~mins for a duration of 1--2~mins depending on the atmospheric conditions. Pointing checks were also made on the phase calibrator every $\sim60$--90~mins. We estimate the absolute flux calibration to be accurate within $\sim$20\%. All ATCA observational and calibration details are summarized in Table~\ref{table_obs}.

The data were calibrated using the MIRIAD package \citep{1995ASPC...77..433S}. We image the data observed at 30 August 2014 at 33/35 GHz with the CLEAN task in MIRIAD using natural weighting, which resulted in a restored beam of $0.43\arcsec \times 0.41\arcsec$ at PA = 72\degr. The size and flux of  HD~97048 for this dataset were determined using the MIRIAD task \textit{uvfit} to fit a Gaussian to the visibility data. The relatively poor seeing during the 31 August 2014 synthesis track, coupled with the higher frequency setting, meant that the quality of the phase correction was significantly worse than on the previous day. Phase decorrelation could be seen in the visibilities, and consequent probable phase errors -- such as radial extensions -- in the reconstructed images. We therefore do not present the image here, rather only the flux determined during the period of best seeing. 

We extract the fluxes from these observations using various methods, including standard model fitting (Point, Gaussian or Point+Gaussian) with MIRIAD's \textit{uvfit} task. However, since \textit{uvfit} does not produce a convenient `goodness-of-fit' criterion, we also use the Miriad task \textit{uvaver} to average all 2048 channels, and then output the resulting uv dataset as a FITS file using the \textit{fits} task. This was subsequently treated in IDL, where the channel-averaged and binned visibilities were fitted with a Point, Gaussian or Point+Gaussian model. The model was then extrapolated to zero spacing (i.e. 0 k$\lambda$) to provide a `predicted' total flux. 
A note is warranted concerning the 5.5 and 9~GHz data, and especially the latter, which was affected by radio frequency interference and other noise that appeared electronic in origin, revealed as an RFI-like (but lower level) pattern on every fourth channel (i.e. a 4 MHz 'period') across particular ranges. All affected channels were removed. 

All the fluxes and, where relevant, the source sizes, are summarized in Table \ref{tab:flux}, where multiple observations at particular frequencies have been averaged to provide a mean and standard deviation. Pre-empting the discussion to follow, it was found that the 33/35, 38/40 and 43/45~GHz data were well fit with a model comprising an unresolved (point) source and a Gaussian, centred at the same position. The fluxes listed in Table \ref{tab:flux} are the total integrated flux.

\begin{figure*}
   \centering
   \includegraphics[width=\hsize]{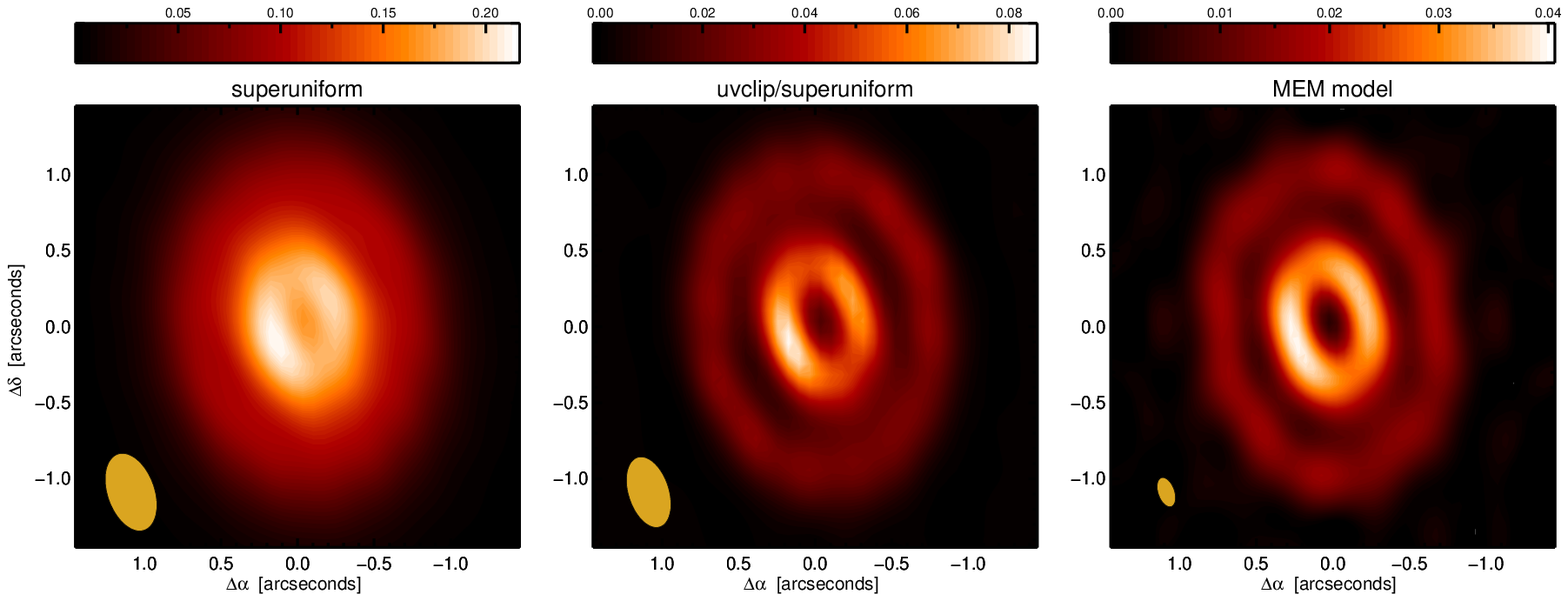}
     \caption{Reconstructed images of \object{HD~97048} for the ALMA band 7 data. The intensity scale for all images is in units of Jy beam$^{-1}$. To highlight the small-scale structure we show (from left to right) images reconstructed with progressively smaller beams. The left panel shows the image reconstructed using superuniform weighting. The central panel shows the superuniformly weighted reconstructed image resulting from clipping all baselines below 160 k$\lambda$. Finally, the right panel shows the MEM model image. The difference in dynamic range between the panels is due to flux loss incurred by clipping the shortest baselines for the central panel, and due to the smaller reconstructed beam for the right panel.} 
    \label{fig:continuum_cycle2}
\end{figure*}

\subsection{ALMA cycle 2 band 7 data}
\label{sec:observations_cycle2}

ALMA Early Science Cycle 2 observations were conducted on 22 May 2015 with 430 seconds of total time on HD~97048 (PI: G van der Plas, Program ID 2013.1.00658.S). The array configuration provided baselines ranging between 21.4 and 555.5 meters. During the observations the perceptible water vapor in the atmosphere varied between 0.62 and 0.82~mm with a median value at zenith of 0.687 mm. 
Two of the four spectral windows of the ALMA correlator were configured in Time Division Mode (TDM) to maximise the sensitivity for continuum observations (128 channels over 1.875~GHz usable bandwidth). These two TDM spectral windows were centred st 345.8~GHz and 356.7~GHz. The other two spectral window were configured in Frequency Division Mode (FDM) to target the  \element[ ][12]{CO} J=3-2 and the \element[+][]{HCO} J=4-3 lines with a spectral resolution of 105 and 103 {\gc m} s$^{-1}$ respectively, using 0.23~GHz of total bandwidth. The data were calibrated using the \textit{Common Astronomy Software Applications} pipeline \citep[CASA, ][version 4.3]{2007ASPC..376..127M}, after which some additional flagging was performed.  Details of the observations and calibration are summarised in Table \ref{table_obs}. We estimate the absolute flux calibration to be accurate within $\sim$10\%. 

We used the standard reduction tools within the CASA package to calibrate and combine the data. We extract the size and flux of HD~97048 by using the CASA task \textit{uvmodelfit} to fit a disk to the visibility data, while we image the disks with the \textit{CLEAN} task in CASA \citep{1974AAS...15..417H} using Briggs (robust = 0.5), uniform and superuniform weighting, which results in a restoring beam of respectively  0.63\arcsec $\times$ 0.36\arcsec at PA = 19\degr (Briggs),  0.60\arcsec $\times$ 0.32\arcsec at PA = 20\degr (uniform) and 0.53\arcsec $\times$ 0.31\arcsec at PA = 19\degr (superuniform). The resulting images are strongly limited dynamically by the bright continuum source and we perform self-calibration on both the phase and amplitude, resulting in a final dynamical range of $\approx$ 1200 and a RMS of 0.20 mJy beam$^{-1}$.  We shift all images using the proper motion correction found in \citet{2007A&A...474..653V} and these are  shown in Figs. \ref{fig:continuum_all} and \ref{fig:continuum_cycle2} for the continuum emission. We applied the self-calibration solutions obtained from the continuum emission to the \element[+][]{HCO} J=4-3 and \element[ ][12]{CO} J=3-2 data and subtracted the continuum emission using the CASA task \textit{uvcontsub}, resulting in a per-channel RMS of 19.5 and 23.7 mJy beam$^{-1}$ respectively. The resulting moment 0, 1 and 8 maps and spectra are shown in Figs. \ref{fig:moments-co32} and \ref{fig:moments-hcop} for the \element[ ][12]{CO} J=3-2 and \element[+][]{HCO} J=4-3.

\subsubsection{Imaging the long baselines}  We also reconstructed the images using only baselines above 160 k$\lambda$ to enhance the contrast of faint, small-scale structure in the disk. These data were cleaned using superuniform weighting, resulting in a beam of 0.48\arcsec $\times$ 0.26\arcsec at PA = 18\degr. The reconstructed image is shown in the central panel of Fig. \ref{fig:continuum_cycle2}.

\subsubsection{MEM image reconstruction} A non-parametric least-squared modelling technique for image reconstruction was performed on the band 7 data. This Maximum Entropy Method (MEM) yields an image with a smaller beam and traces finer spatial scales than the previously described CLEANed reconstruction. Examples of usage of MEM for image synthesis in astronomy can be found in e.g. \citet{1978Natur.272..686G, 2015ApJ...813...76M}. Images deconvoled with MEM "super-resolve" the interferometric data, as the entropy prior allows an extrapolation of spatial frequencies beyond those sampled by the interferometer. We use the \textit{uvmem} algorithm \citep{2006ApJ...639..951C, 2015ApJ...811...92C, 2015ApJ...812..126C} and  label the resulting model as "MEM model" in the right panel of Fig. \ref{fig:continuum_cycle2}. The spatial resolution reached in this reconstruction is about 1/3 the clean beam calculated with uniform weights. 

\subsection{ALMA Cycle 1 band 3 data}
\label{sec:observations_cycle0}

ALMA Early Science Cycle 1 observations were conducted on 1 December  2013 with 72.6 seconds of total time on HD~97048 \citep[Program ID 2012.1.00031.S, see also][]{2016ApJ...823..160D}. The array configuration provided baselines ranging between 15.8 and 462.9 meters. During the observations the perceptible water vapor in the atmosphere were stable within 5\% of the median value of 0.535 mm. Three of the four spectral windows of the ALMA correlator were configured in TDM to maximise the sensitivity for continuum observations (128 channels over 1.875~GHz usable bandwidth). These spectral windows were centred at 101.9 ~GHz, 103.9~GHz and 113.1~GHz. The fourth spectral window was configured in FDM to target the  \element[ ][12]{CO} J=1-0 line and centered at 105.2~GHz, with a spectral resolution of 159 m s$^{-1}$ and a total bandwidth of 0.117~GHz. The data were calibrated using the provided data reduction script and CASA version 4.3.  Upon inspecting the visibilities, the amplitudes of the spectral window centered at 113.1~GHz showed anomalous behaviour and we decided to flag the entire spectral window. We thus only use the first two spectral windows for the dust continuum analysis, and estimate the absolute flux calibration to be accurate within $\sim$10\%. Details of the observations and calibration are summarised in Table \ref{table_obs}.

We used the same data reduction process using CASA as for the band 7 data.  After CLEANing the data using uniform weighting we obtain a restoring beam of 2.36\arcsec $\times$ 1.40\arcsec at PA = -39\degr.  Self-calibration on both the phases and the amplitude results in a continuum RMS of 0.18 mJy beam$^{-1}$. The continuum emission is shown in the central panel of Fig. \ref{fig:continuum_all}. Again we applied the self-calibration solutions obtained from the continuum emission to the \element[ ][12]{CO} data, and after subtracting the continuum, the line data were imaged using Briggs weighting (robust = 0.5). This  resulted in a restoring beam of 2.21\arcsec $\times$ 1.43\arcsec at PA = -38\degr and a RMS of 77.5 mJy beam$^{-1}$. The \element[ ][12]{CO} J=1-0 moment 0, 1 and 8 map and line spectrum are shown in Fig. \ref{fig:moments-co10}. 

\section{Results} \label{sec:results}

\subsection{Continuum emission}

We detect the disk around HD~97048 in the ALMA band 3 and band 7 observations, as well as in the ATCA observations. We spatially resolve the continuum emission in the ALMA band 3, ALMA band 7, and ATCA 33/35, 38/40, 43/45~GHz and 93/95 extended array observations, and even the 91/97~GHz hybrid configurations. While the 3~mm to 7~mm ATCA observations are resolved and we fit the continuum emission with a Gaussian to derive their fluxes, we only use the 33/35 GHz data for imaging. We list the extracted flux densities and the deconvolved size, inclination and position angle derived by fitting a disk or Gaussian in the uv plane in Table \ref{tab:flux} and show the CLEANed continuum maps in Fig. \ref{fig:continuum_all}. We adopt the disk inclination of 41.4 $\pm$ 0.9\degr and the position angle of 4.5 $\pm$ 0.1\degr, as determined from the ALMA band 7 continuum observations, throughout this work. 

A disk cavity is directly visible in the ALMA band 7 image, but surprisingly is not detected in the other observations (e.g. ATCA 29 August 2014) despite the spatial resolution being adequate. The ALMA band 7 image also shows a break in the radial brightness distribution at a distance of $\approx$ 1\arcsec (158 au). We highlight this structure in Fig. \ref{fig:continuum_cycle2}, where we show the image CLEANed using super-uniform weighting, an image reconstructed using only the baselines $>$ 160 k$\lambda$ to emphasize faint structures, and an image reconstructed using MEM rather than to CLEAN. Both the inner cavity and the break can be seen in the integrated radial surface {\gc brightness} profile as shown in Fig. \ref{fig:radial_intensity}, where we show the superuniformly CLEANed ALMA band 7 image deprojected using the disk position angle and inclination listed in Table \ref{tab:flux}, in the bottom right panel. The same image transformed to polar coordinates is shown in the top left panel and is collapsed in the radial {\gc (top right panel) and azimuthal (bottom left panel)} dimension to obtain the respective integrated surface brightness structures. Radially, the intensity peaks at $\approx$ 0.3\arcsec, while there is a `shoulder' present in the intensity profile between $\approx$ 0.8 and 1.2\arcsec. Ultimately, another break in the slope of the surface brightness is visible at a distance of 1.90\arcsec.

The radial surface brightness distribution of the disk outside of the cavity can be approximately described by 3 power laws, as is best seen on a logarithmic scale shown in the inset in the bottom left panel of Fig. \ref{fig:radial_intensity}.  We fit a power law to these 3 sections of the form \textit{a$_0$ $\times$ r$^{a_1}$} between 0.3 and 2.6\arcsec, separated at 1.05\arcsec and 1.95\arcsec. This results in best fit values for [a$_0$, a$_1$] {\gc to the normalized radial surface brightness} of [0.5, -0.8], [0.7, -5.7] and [14.3, -10.4], respectively. Azimuthally the brightness fluctuates with an amplitude of at most 4\%, most of which can be accounted for by image reconstruction artifacts (see \S \ref{disc:mcfost}). 

The ALMA band 3 image is spatially resolved but shows neither an inner cavity nor a radial structure. 
The 34~GHz ATCA image, finally, shows a peak of emission coincident with the stellar position after correction for its proper motion. 

The SED between 0.853~mm and 5.45~cm can be constructed for HD~97048 using the ATCA and ALMA fluxes reported here, complemented by 1.3~mm SEST fluxes \citep{1993A&A...276..129H, 1998A&A...336..565H} and 5.5 to 19~GHz re-reduced archival ATCA observations. Some of the latter have previously been presented in \citet{CUBACH-phd}. 
The HD~97048 SED is shown in the left panel of Fig.~\ref{fig:SED}. A break at $\approx$ 10--13~mm separates two parts which are probably  dominated by different emission mechanisms. At shorter wavelengths,  thermal dust emission from the disk dominates, whilst the flattening of the SED at longer wavelengths suggests a different emission mechanism is responsible for much of that emission. This could feasibly be thermal wind (free-free) emission from a stellar, a disk outflow {\gc or a photo-evaporating disk \citep{2012ApJ...751L..42P}}, or even a non-thermal process from the star itself or the star-inner disk boundary.

A linear least squares fit to the 1--10~mm portion of the SED yields a spectral index $\alpha$ (where $F \propto \nu^\alpha$) of 3.1 $\pm$ 0.1. 
The spectral slope beyond 10~mm is more difficult to reliably determine for several reasons. For instance, the S/N is lower (due to only 20--60~minute on-source times), the source is much fainter at longer wavelengths, and the data is effected by RFI. The lower spatial resolution of the ATCA hybrid arrays combined with the increased `background' of extragalactic radio sources makes unique identification of HD~97048 more challenging. Despite these difficulties, we estimate that the fluxes at 5.5 and 9~GHz are approximately the same at about 0.3~mJy, but with a factor of 2 uncertainty. To obtain a spectral index for the measurements beyond 10~mm, we first subtract the extrapolated contribution of the power-law obtained from the measurements between 1 and 10~mm, and obtain a best fit to these corrected measurements of $\alpha$ = 0.4 $\pm$ 0.6.
We discuss the ATCA data in more detail in \S \ref{disc:free}.

\subsection{CO and \element[+][]{HCO} emission}

\begin{figure*}
   \centering
   \includegraphics[width=\hsize]{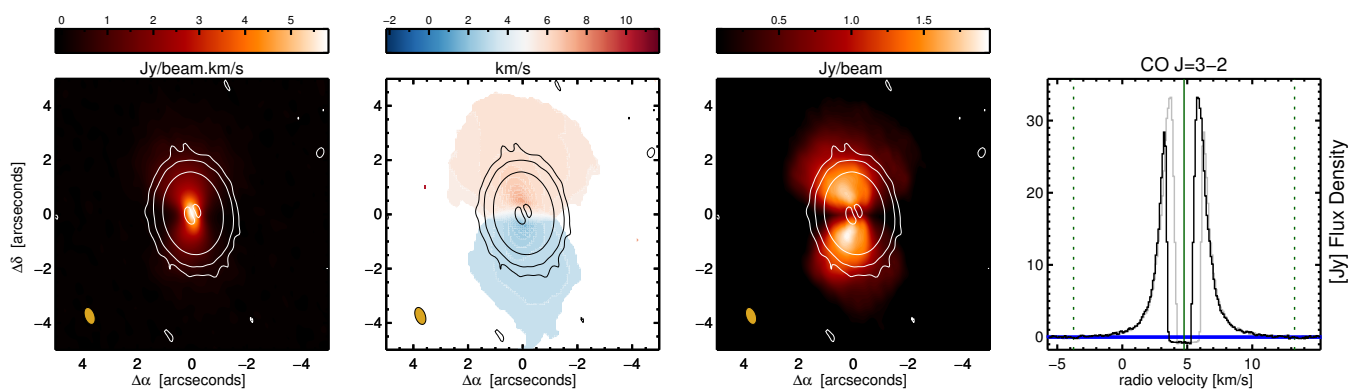}
     \caption{Summary of \element[ ][12]{CO} J=3-2 line emission in \object{HD~97048}. We show the integrated intensity (moment 0, left panel), intensity-weighted velocity  (moment 1, 2$^{\mathrm{nd}}$ panel), peak intensity (moment 8, 3$^{\mathrm{rd}}$ panel) and the collapsed emission line (right panel). Each moment map was made using a 3 $\sigma$ cutoff and emission imaged using briggs weighting. Over plotted in the three left panels are the same continuum contours as in the left panel of Fig. \ref{fig:continuum_all}. The beam is shown in orange in the bottom left of each panel. In the right panel we show the spectrum mirrored at the systemic velocity with a grey line.  The grey shaded area denotes the + and - 3$\sigma$ level calculated outside the line boundaries.}
    \label{fig:moments-co32}
\end{figure*}

\begin{figure*}
   \centering
   \includegraphics[width=\hsize]{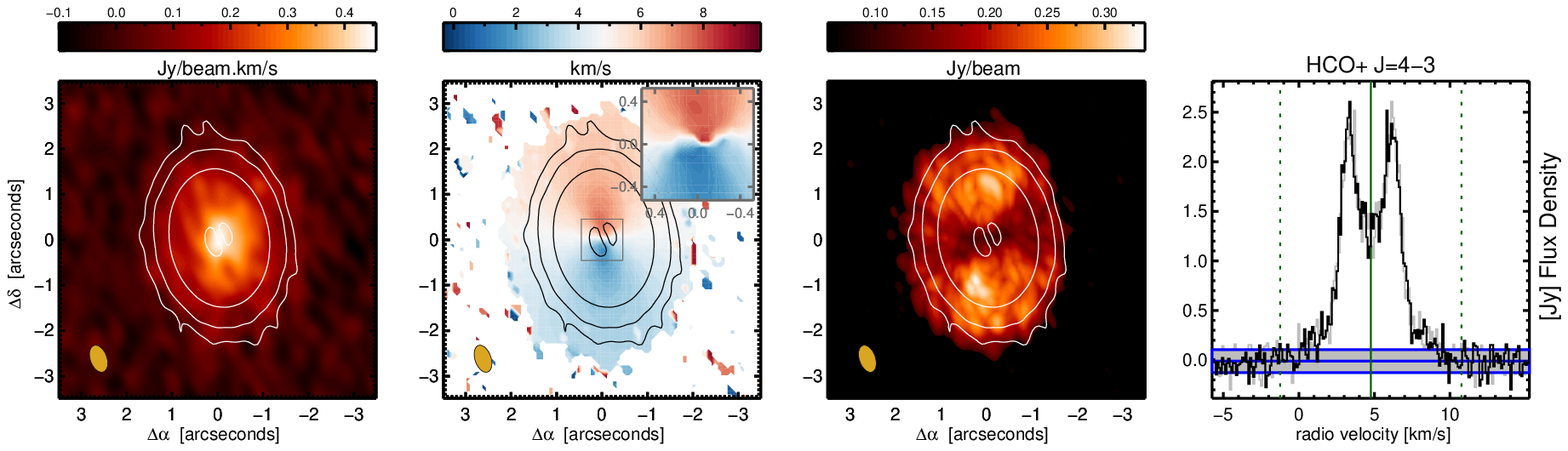}
     \caption{Summary of \element[+][]{HCO} J=4-3 line emission in \object{HD 97048}. We show the integrated intensity (moment 0, left panel), intensity-weighted velocity  (moment 1, 2$^{\mathrm{nd}}$ panel), peak intensity (moment 8, 3$^{\mathrm{rd}}$ panel) and the collapsed emission line (right panel). Each moment map was made using a 3 $\sigma$ cutoff and emission imaged using briggs weighting. Over plotted in the three left panels are the same continuum contours as in the left panel of Fig. \ref{fig:continuum_all}, and the inset at the velocity map shows the inner 0.5\arcsec of the velocity map made using a 5 $\sigma$ cutoff. The beam is shown in orange in the bottom left of each panel. In the right panel we show the spectrum mirrored at the systemic velocity with a grey line. The grey shaded area denotes the + and - 3$\sigma$ level calculated outside the line boundaries.}
    \label{fig:moments-hcop}
\end{figure*}

\begin{figure*}
   \centering
   \includegraphics[width=\hsize]{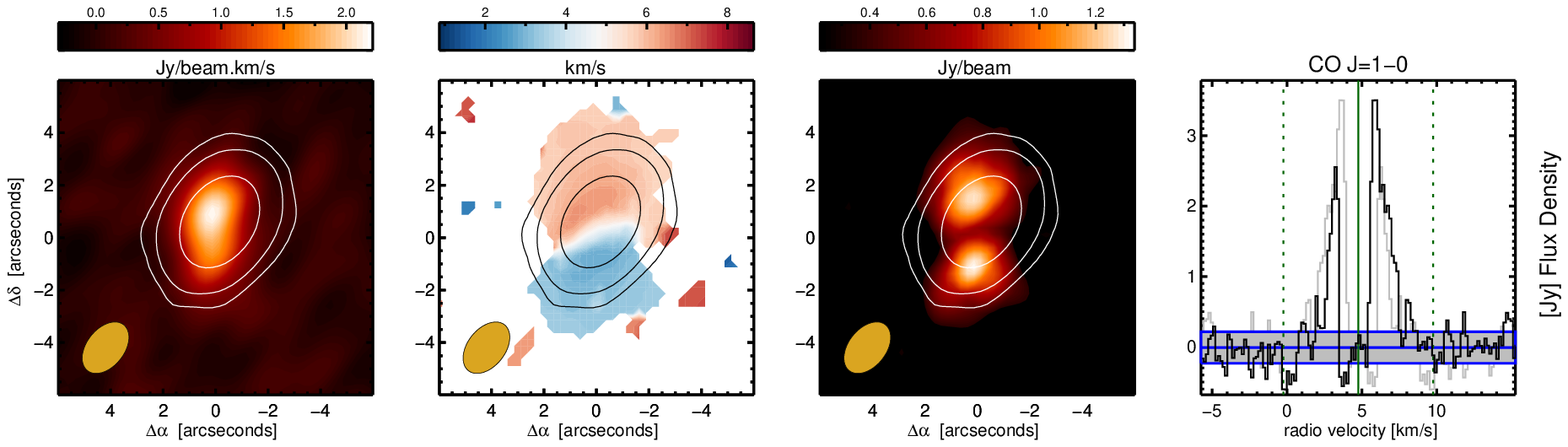}
     \caption{Summary of \element[ ][12]{CO} J=1-0 line emission in \object{HD~97048}. We show the integrated intensity (moment 0, left panel), intensity-weighted velocity  (moment 1, 2$^{\mathrm{nd}}$ panel), peak intensity (moment 8, 3$^{\mathrm{rd}}$ panel) and the collapsed emission line (right panel). Each moment map was made using a 3 $\sigma$ cutoff and emission imaged using briggs weighting. Over plotted in the three left panels are the same continuum contours as in the central panel of Fig. \ref{fig:continuum_all}. The beam is shown in orange in the bottom left of each panel. In the right panel we show the spectrum mirrored at the systemic velocity with a grey line. The grey shaded area denotes the + and - 3$\sigma$ level calculated outside the line boundaries.}
    \label{fig:moments-co10}
\end{figure*}

We spatially and spectrally resolve the \element[ ][12]{CO} J = 1-0, \element[ ][12]{CO} J = 3-2 and \element[+][]{HCO} J = 4-3 emission in the HD~97048 disk. The integrated intensity (moment 0), intensity-weighted velocity  (moment 1) and peak intensity (moment 8) maps, plus the spectra, are shown in Fig. \ref{fig:moments-co32}, \ref{fig:moments-hcop} and \ref{fig:moments-co10} for the \element[ ][12]{CO} J=3-2 emission, the \element[+][]{HCO} J=4-3 emission, and  the \element[ ][12]{CO} J=1-0 emission respectively. The integrated fluxes and disk sizes as measured along the disk major axis for all line emission $>$ 3$\sigma$ are listed in Table \ref{tab:flux_lines}.

The \element[ ][12]{CO} emission shows a broad absorption feature between 3.4 and 5.7 km s$^{-1}$, completely blocking line and continuum emission. The drop below 0 in the spectrum seen most clearly in the right panel of Fig. \ref{fig:moments-co32} is a natural consequence of subtracting the continuum over the channels where  both the continuum and line emission are filtered out by foreground {\gc absorption}. The \element[+][]{HCO} line appears unaffected by this foreground absorption. 

We determine the systemic velocity from the \element[+][]{HCO} spectrum to be v$_{LSR}$ = 4.75 $\pm$ 0.1 km s$^{-1}$ based on both the peak-to-peak separation and on the symmetry of the emission in the channel maps, {\gc of which we show a subset in Fig. \ref{fig:channels}. We show the complete set of channel maps in the Appendix (Figs. \ref{fig:channels-co-1} to \ref{fig:channels-hco-2})}.  The \element[ ][12]{CO} and \element[+][]{HCO} integrated emission maps peak inside of the disk cavity. The \element[ ][12]{CO} J=3-2 and \element[+][]{HCO} J=4-3 lines have a maximum half width in the line wings of respectively 8.5 and 6.0 km s$^{-1}$, determined from the presence of emission $>$ 3 $\sigma$ in the channel maps. These values can be translated to an emitting radius of 13.4 and 27.0 au, assuming the gas is in Keplerian rotation in a disk inclined by 41.4\degr around a 2.5 M$_{\odot}$ star.
The gas disk as measured for all three lines is more extended than the continuum emission. The outer radius measured along the semi-major axis, considering only the emission above 3 $\sigma$ and deconvolved with the beam for the \element[ ][12]{CO} J=1-0, J=3-2 and \element[+][]{HCO} J=4-3, is 4.2\arcsec, 5.2\arcsec and 2.9\arcsec,  respectively. This is a factor of 1.9, 2.3 and 1.3 larger than the continuum emission.

\begin{table*}
\setlength{\tabcolsep}{6pt} 

\caption{Continuum fluxes, major axis, axis ratio and position angle derived from fitting a disk (`D', for the ALMA data) or a Gaussian or point source (`G' or `P', for the ATCA data) to the visibilities. The flux error values do not include errors on flux calibration, which we estimate to be 10\% for the ALMA observations and 20\% for the ATCA observations. The 5.5, 9, 17 and 19 GHz ATCA fluxes were earlier published in \citet{CUBACH-phd}. We re-extract the fluxes from the observations following the same method used for the other ATCA fluxes for consistency.}\label{tab:flux} 
\smallskip
\centering   
\tiny
\noindent\begin{tabularx}{\textwidth}{@{\extracolsep{\stretch{1}}}*{8}{l}@{}}

\hline

ID & wavelength & flux & error & uv model & major axis & inclination & PA \\
            & mm         & mJy   & mJy  & & arcsec     &  degrees    & degrees \\
\hline
ALMA band 7 & 0.853    & 2253.16 & 0.96 & D & 2.25 (0.01) & 41.4$^{\pm 0.9}$ & 4.5 (0.1) \\
ALMA band 3 & 2.939    & 92.0  & 2.5 & D & 2.27 (0.02) & 40.5$^{\pm 0.9}$  & 2.3 (1.7) \\

\hline

ATCA 97 GHz & 3.091  & 76.3  & 5.7  & G  & -  &  - &  - \\
ATCA 95 GHz & 3.156  & 62.3  & 4.7  & G  & -  &  - &  - \\
ATCA 93 GHz & 3.224  & 63.0  & 3.5  & G  & -  &  - &  - \\
ATCA 91 GHz & 3.294  & 62.9  & 3.0  & G  & -  &  - &  - \\
ATCA 45 GHz & 6.662  & 5.2   & 0.7  & G  & -  &  - &  - \\
ATCA 43 GHz & 6.972  & 5.1   & 0.5  & G  & -  &  - &  - \\
ATCA 40 GHz & 7.459  & 2.67  & 0.21 & G  & -  &  - &  - \\
ATCA 38 GHz & 7.889  & 2.36  & 0.23 & G  & -  &  - &  - \\
ATCA 35 GHz & 8.565  & 2.51  & 0.18 & G  & 1.7  & 49.5 & -16  \\
ATCA 33 GHz & 9.085  & 2.07  & 0.25 & G  & 1.6  & 55.9 & -16  \\
ATCA 24 GHz & 12.491 & 1.16  & 0.34 & P  & -  &  - &  - \\
ATCA 19 GHz & 15.779 & 0.80  & 0.09 & P  & -  &  - &  - \\
ATCA 18 GHz & 16.655 & 0.72  & 0.16 & P  & -  &  - &  - \\
ATCA 17 GHz & 17.635 & 0.70  & 0.06 & P  & -  &  - &  - \\
ATCA 9 GHz  & 33.310 & 0.3  & $^{+0.3}$ $_{-0.15}$  & P  & -  &  - &  - \\
ATCA 5.5 GHz & 54.508 & 0.3 & $^{+0.3}$ $_{-0.15}$ & P  & -  &  - &  - \\

\hline                                   
\end{tabularx}

\end{table*}
\begin{table}
\setlength{\tabcolsep}{4pt} 

\caption{Line fluxes, spectral resolution and spatial extent for the CO J=1-0, J=3-2 and HCO$^+$ J = 4-3 lines.\label{tab:flux_lines}} 
\smallskip
\centering   
\scriptsize
\noindent\begin{tabularx}{\columnwidth}{@{\extracolsep{\stretch{1}}}*{6}{l}@{}}
\hline

Line  & line flux$^{\mathrm{a}}$ & error$^{\mathrm{b}}$ & Channel width  & RMS$^{\mathrm{c}}$   & major axis$^{\mathrm{d}}$  \\
      & Jy        &  Jy   & m s$^{-1}$  & mJy beam$^{-1}$          &  arcseconds \\
\hline
CO J = 1-0      &  8.22  & 0.28 & 159 & 77.5 & 4.2 \\
CO J = 3-2      &  74.28 & 0.14 & 106 & 23.7 & 5.2 \\
HCO$^+$ J = 4-3 &  9.44  & 0.13 & 103 & 19.5 & 2.9 \\
\hline                                   
\end{tabularx}
\tablefoot{\textbf{$^{\mathrm{a}}$:} The line flux has been integrated only over the channels with positive flux, and is a lower limit for both CO lines due to the foreground extinction in the line core. \textbf{$^{\mathrm{b}}$:} Estimated from the RMS of the spectrum outside the line boundaries, does not include calibration uncertainties. \textbf{$^{\mathrm{c}}$:} 1 $\sigma$ RMS per channel, determined from all channels outside the line boundaries. \textbf{$^{\mathrm{d}}$:} The major axis is determined for all emission above 3 times the RMS per channel and has been deconvolved from the beam.}
\end{table}

\begin{figure*}
   \centering
   \includegraphics[width=\textwidth]{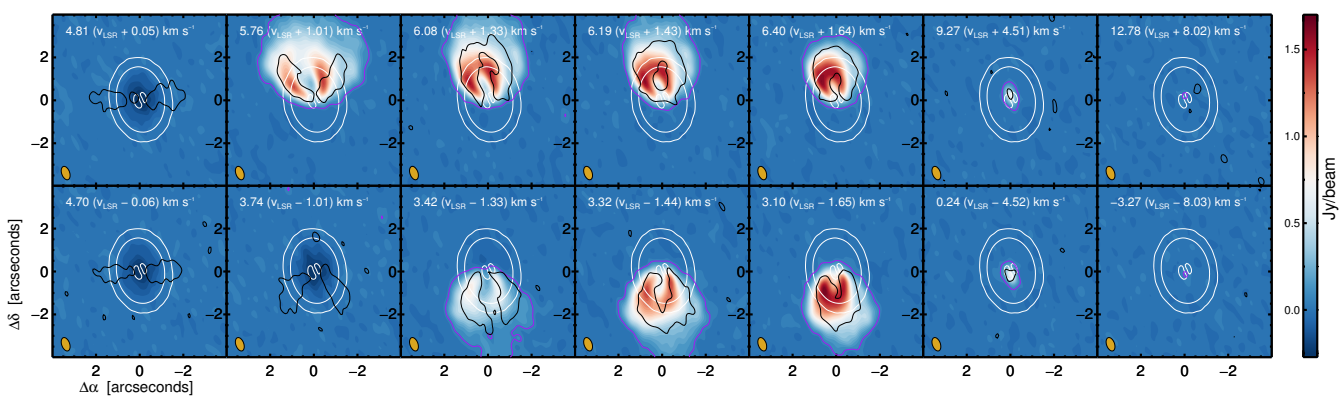}
     \caption{Selected channel maps showing the \element[ ][12]{CO} channels in color and the 3 $\sigma$ outline in purple, the \element[+][]{HCO} emission 3 $\sigma$ outline with black contours, and the continuum contours in white. In the top right of each panel we note the v$_{lsr}$ in white with the velocity with respect to the systemic velocity of 4.75 km s$^{-1}$ in parenthesis. The clean beam is shown in orange in the bottom left of each panel. Note the foreground absorption for the CO emission in the leftmost panels, and the emission inside the disk gap in the rightmost panels. {\gc Complete channel maps are shown in Figs. \ref{fig:channels-co-1} to \ref{fig:channels-hco-2} in the Appendix.}} 
    \label{fig:channels}
\end{figure*}

\begin{figure}
   \centering
   \includegraphics[width=\hsize]{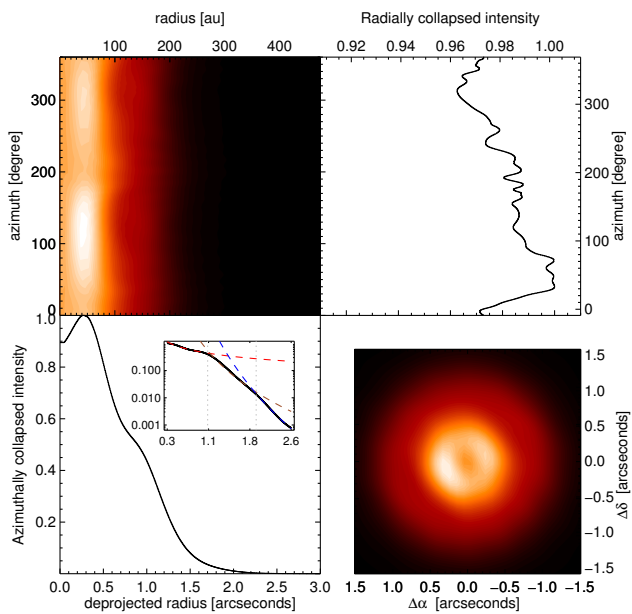}
     \caption{ALMA Band 7 image reconstructed using superuniform weighting as described in \S \ref{sec:observations_cycle2}, deprojected using the disk inclination and position angle listed in Table \ref{tab:flux}, bottom right panel, and converted to polar coordinates (top left panel). This map is collapsed along the radial and azimuthal axes to yield the azimuthal intensity distribution (top right) and the radial intensity distribution (bottom left). With an inset in this last panel we zoom in between 0.3 and 2.6\arcsec with the normalized intensity plotted in log scale. We also overplot the 3 power law fits discussed in \S \ref{sec:results}, separated at 1.05 and 1.95\arcsec.}
    \label{fig:radial_intensity}
\end{figure}

\begin{figure}
   \centering
   \includegraphics[width=\hsize]{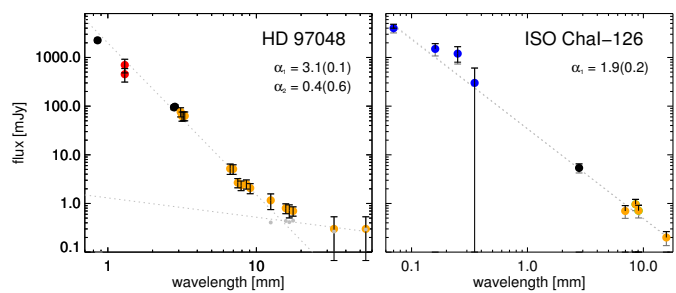}
     \caption{SED for HD~97048 (left panel) and ISO ChaI-126 (right panel). The error bars shown in both panels include a calibration uncertainty of 10\% (ALMA), 20\% (ATCA + Herschel) and 30\% (SEST). The \textbf{left panel} shows a compilation of all points beyond 850 $\mu$m including literature data from SEST \citep[red, ][]{1993A&A...276..129H, 1998A&A...336..565H}, ALMA (black) and ATCA (orange) data presented in this paper. Overplotted with grey lines are linear least squares fits to parts of the SED between 1 and 10~mm and between 10~mm and 5.45~cm. The latter fit was made using fluxes corrected with extrapolated fluxes from the 1-10~mm fit. These corrected fluxes are shown with small grey dots.  The \textbf{right panel} shows a compilation of all points beyond 70 $\mu$m including literature data from Herschel \citep[blue, ][]{2012A&A...545A.145W}, ALMA (black) and ATCA (orange) data presented in this paper. Overplotted with grey lines is a linear least squares fit to all data points.}
    \label{fig:SED}
\end{figure}

\subsection{ISO-ChaI~126} \label{res:iso-cha1-126}

\begin{figure}
   \centering
   \includegraphics[width=\hsize]{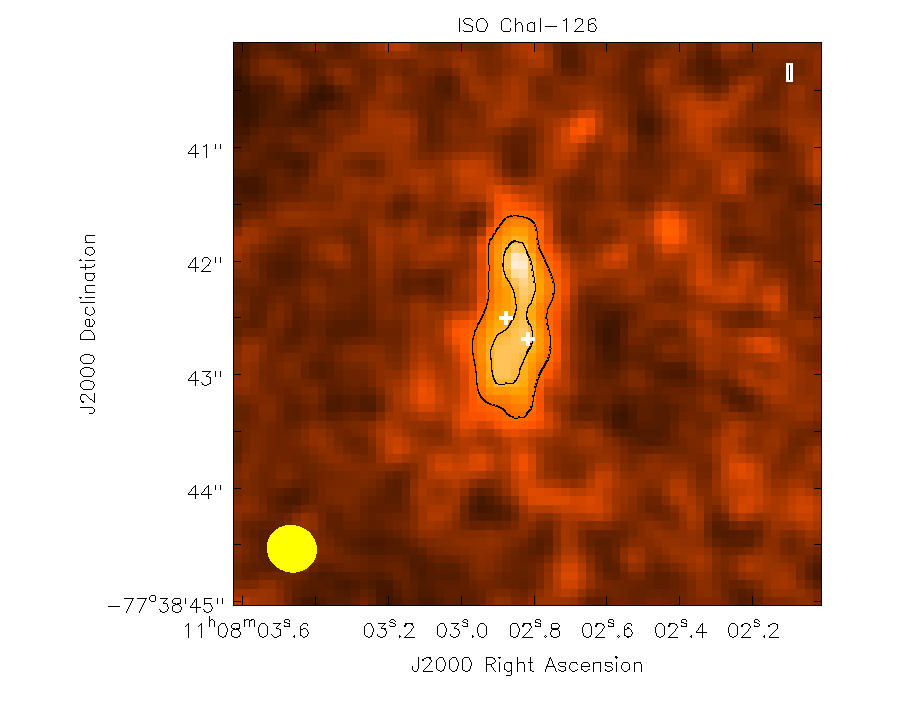}
     \caption{34~GHz emission from the ISO-Cha1~126 binary system resolved with ATCA.  Overplotted with black contours we show 5 and 7.5 times the RMS of 0.010 mJy beam$^{-1}$, the beam is shown in yellow in the bottom left corner. The white `+' signs denotes the J2000 coordinates of the A and B components corrected for the proper motion of HD~97048.}
    \label{fig:iso-cha1-126}
\end{figure}

ISO-ChaI~126 is a $0.5 M_{\odot} + 0.5 M_{\odot}$ binary, located 35\arcsec north of HD~97048 and part of the Chameleon I star forming region. It is resolved using NACO K$_s$ band imaging by \citet{2013A&A...554A..43D}, with A and B components separated by 0.28\arcsec at a position angle of 228.8\degr. We detect emission from the ISO-ChaI~126 system in both the ALMA band 3 and the ATCA data at 18, 34 and 44~GHz. The system is spatially resolved only in the ATCA observations at 34 GHz, which we show imaged using natural weighting in Fig. \ref{fig:iso-cha1-126}. We list the integrated flux values, derived from fitting a Gaussian to the visibilities, in Table \ref{tab:flux-126}.  We resolve an elongated structure in the 34~GHz ATCA data with a total flux of 0.7 $\pm$ 0.1 mJy, and a major and minor  axis of 1.89 $\pm$ 0.24\arcsec and 0.62 $\pm$ 0.09\arcsec at a position angle of 174 $\pm$ 4\degr deconvolved from the beam when fitting a Gaussian to the image. The stellar positions fall approximately in the center of this structure when corrected for the proper motion of HD~97048.

\begin{table}
\setlength{\tabcolsep}{4pt} 

\caption{Continuum fluxes for ISO ChaI-126 determined from fitting a Gaussian to the uv plane. The flux error values do not include errors on flux calibration, which we estimate to be 10\% for the ALMA observations and 20\% for the ATCA observations.}\label{tab:flux-126}.  
\smallskip
\centering   
\scriptsize
\noindent\begin{tabularx}{\columnwidth}{@{\extracolsep{\stretch{1}}}*{4}{l}@{}}
\hline

ID & wavelength & flux & error \\
            & mm         & mJy   & mJy   \\
            \hline
ALMA band 3       & 2.939     & 5.4     & 1.0 \\
ATCA 43 GHz       & 6.972     & 0.70    & 0.15 \\
ATCA 35 GHz       & 8.565     & 0.71    & 0.15 \\
ATCA 33 GHz       & 9.085     & 0.96    & 0.16 \\
ATCA 19 GHz       & 15.779    & 0.20    & 0.05 \\

\hline                                   
\end{tabularx}

\end{table}

Linear least squares fits to the ALMA and ATCA detections, complemented by Herschel detections reported by \citet{2012A&A...545A.145W}, yield a spectral index $\alpha$ = 1.9 $\pm$ 0.2 (Fig. \ref{fig:SED}, right panel). We discuss our interpretation of ISO-ChaI~126 in Sect. \ref{dis:iso-cha1-126}.

\section{Discussion}\label{sec:discussion}

The data presented here indicate at face value that the continuum emission from the disk around HD~97048 extends out to 2.25\arcsec (355 au) and shows radial structure including a drop in intensity out to $\approx$ 0.3\arcsec and a `shoulder' further out at $\approx$ 1\arcsec. The gas disk, as most accurately traced by the \element[ ][12]{CO} J=3-2 emission, extends up to a factor of 2.4 beyond the continuum emission. 

The radial brightness distribution of the dust continuum emission between 0.3 and 2.6\arcsec can be loosely fitted by 3 power law fits separated at 0.95\arcsec and 1.95\arcsec with indices of -0.8, -5.7 and -10.4 respectively. Such a 3-stage power law with increasingly smaller indices towards the outer disk reflect a reasonable representation for the radial dust distribution when solving the bulk transport of solids embedded in a gas-rich, viscous accretion disk. In this case, the first slope represents a fragmentation-limited dust distribution in the inner disk, the second a drift-dominated dust distribution in the outer disk, and the last a steep drop outside of the disk outer edge \citep{2014ApJ...780..153B}. We thus interpret the final part of the radial intensity distribution between 1.95\arcsec and 2.6\arcsec as a smearing effect by the beam of a sharp outer disk edge at 2.25\arcsec. The observed steep drop in mm-dust emission and a more extended gas disk is consistent with size sorting of the larger grains through radial drift \citep[e.g.][]{2014ApJ...780..153B, 2014MNRAS.437.3037L}.

The first break in the radial intensity distribution at $\approx$ 1\arcsec~ separates two slopes that are similar to the broken power law values found for the disk around TW Hya by \citet[][-0.53 and -8]{2016A&A...586A..99H}  and \citet[][-0.7 and -6]{2016ApJ...820L..40A}. Such a broken power law structure can be caused either by a decrease in emission inside 1\arcsec or a boost in emission outside 1\arcsec. Both these options are physically motivated: a decrease in emission could be the result of a gap which is carved out by a forming planet \citep{2005ApJ...619.1114W}, MHD effects \citep{2015A&A...574A..68F}, or be the result of rapid grain growth at a condensation front \citep{2015ApJ...806L...7Z}. An extra emitting ring can be caused by fragmentation and drift-dominated dust distributions, but also by other mechanisms, such as an enhancement or pile-up in small dust grains caused by e.g. the sintering of aggregates \citep{2016ApJ...821...82O}, or by a self-induced dust pile-up resulting from the interplay between aerodynamic drag and the growth, fragmentation and migration of grains as described by \citep{2015MNRAS.454L..36G}.

We use the 3D, Monte-Carlo  based,  radiative  transfer code MCFOST \citep{2006A&A...459..797P,2009A&A...498..967P} to simulate an axisymmetric disk which we compare to our observations to test if these 2 proposed scenarios -- a gap or a ring -- can indeed account for the observed feature.

\subsection{Radiative transfer modeling} \label{disc:mcfost}

We use the MCFOST model parameters from a model based on the SED and mid-IR observations of HD~97048  \citep{2007A&A...470..625D}, and update this model with the disk inclination and position angle listed in Table \ref{tab:flux}. To test whether the above 2 scenarios can be reproduced by this model disk, we either carve a Gaussian gap with variable width and depth in it, or add an extra emitting ring with variable width an mass. We then modify the disk inner and outer radius and radial surface density exponent to match our observations. We choose not to include an inner disk between 0.1 and 2 au into our modeling. Such a disk was modeled by \citet{2013A&A...555A..64M} to account for the NIR excess in the SED. The contribution of such an inner disk to the 0.85 mm flux is $\approx$ 1/1000 compared to the contribution of the outer disk, and its inclusion or exclusion will have no impact on our modeling.

After computing the model disk, we use the CASA tasks \textit{simobserve} to convert the image into visibilities, as would be observed with a similar antenna configuration and on-sky properties as for our ALMA band 7 observations. Finally, we CLEAN the visibilities and compare these simulated observations, both directly using the visibilities and in the image plane, with our data. We stress that our MCFOST models are meant to match the radial brightness distribution of the ALMA band 7 data, and not to be a comprehensive model for the disk at other wavelengths.

\begin{figure*}
   \centering
   \includegraphics[width=\hsize]{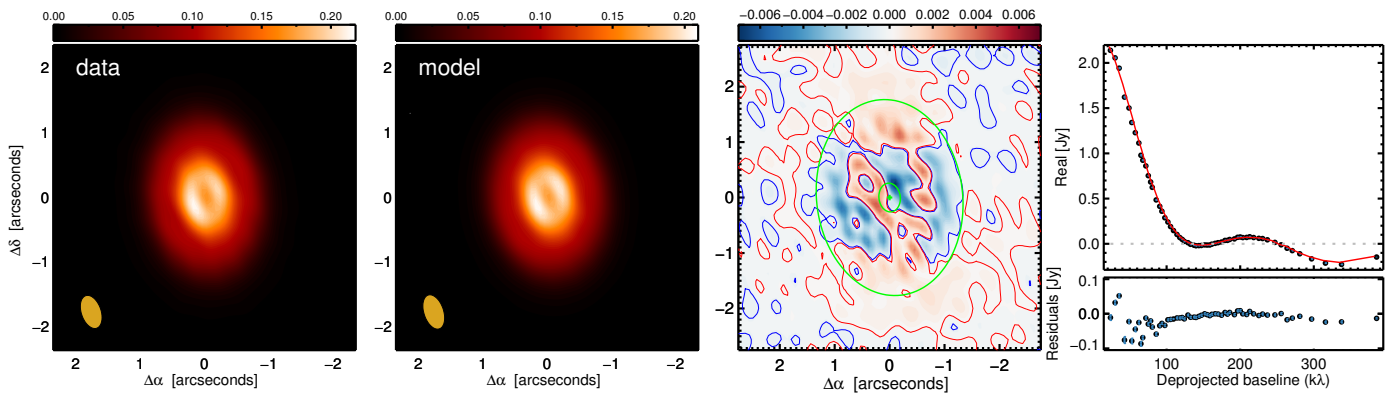}
     \caption{Comparison of ALMA band 7 data (left panel) with the ring model image degraded to the resolution of our data (2$^{\mathrm{nd}}$ panel). The 3$^{\mathrm{rd}}$ panel shows the residuals of the observations minus the model with the inner and outer radius of the model disk at 43.5 and 280 au shown with green ellipses. We draw red and blue contours at $\pm$ 0.2 mJy to highlight the extent of the outer disk in the residual image. Units of all intensity scales are in Jy/beam. The right panel shows the real part of the visibilities as function of the deprojected baseline for the data (top panel, black dots) and the model (red line). The bottom panel shows the residuals. The visibilities are binned to have a constant amount of baselines per bin. Error bars are mostly smaller than the plot symbol size, and are shown with blue vertical lines}
    \label{fig:model_comparison}
\end{figure*}

We find that both tested underlying brightness distributions produce comparable matches to the observations. For example, a sample `gap model’ with a disk between 46 and 290 au with a flat surface density exponent and a gap between 79 and 101 au produces similar residuals to a sample `ring model’ of a continuous disk between 41 and 280 au with a surface density exponent of -1.6 which has extra emission superimposed as a ring between 135 and 180 au. We show the ring model, the residuals and the deprojected visibilities in Fig. \ref{fig:model_comparison}. As our models are meant to explore the possible range of underlying brightness distributions, we do not determine confidence intervals for our derived parameters. Rather, we discuss the general trends learned from our efforts to reproduce the observations.

In general, our observations are well reproduced by a model disk starting between 40 and 46 au and extending out to 280 to 300 au. The width of either the disk gap or the extra emitting ring is not well constrained, as it depends on the depth or intensity of the feature respectively. Radially, the gap is centered at $\approx$ 90 au and the ring further out at $\approx$ 150 au. All our model disks are optically thick at 853 $\mu$m for the first 15 to 20 au starting at the inner rim of the disk. These model disks fit the observations reasonably well with maximum residuals of a few percent of the original image in the inner 200 au. Most notably, these residuals are  present at the location of the inner rim of the disk, where our axisymmetric model under predicts flux at the south-eastern inner rim and over predicts the emission at the north-west side of the inner rim. Furthermore, our best choice for the outer disk radius in the model is about 0.5\arcsec smaller than the size derived from fitting the disk visibilities, which is corroborated by faint emission up to 2.25\arcsec that can be seen in the residual image. Our model disk is truncated sharply at the disk outer edge, but as discussed, the radial intensity distribution shows a gradual but steep decrease beyond 1.95\arcsec. Using a more gradually truncated outer disk prescription such as a tapered exponential will most likely result in a better agreement between our model and the observations.

\subsection{The inner cavity}

The surface brightness of an inclined optically thin disk peaks along the major axis due to projection effects. The ALMA band 7 data in contrast shows a disk whose surface brightness peaks along the disk minor axis. This can be naturally explained by the combination of a disk that is optically thick in the first 10-20 au from the {\gc outer disk inner} rim, such as we find in our models and which negates the projection effect, and the effect of convolution with a beam that is aligned along the major axis of the disk, as is the case for our observations.

In our best models, the disk starts at 40 to 46 au from the star and is truncated sharply at the inner rim radius. Choosing a more gradual transition can shift the inner radius of the disk inwards to between 5 and 10 au. The presence of a disk cavity with a radius of 34$\pm$4 au is predicted by \citet{2013A&A...555A..64M} based on simultaneous modeling of the SED and spatially resolved Q band (20~$\mu$m) spectrum. In contrast, polarimetric differential imaging did not detect any cavity with a size larger than 16 au in radius \citep{2012A&A...538A..92Q}, while spatially resolved \element[ ][12]{CO} ro-vibrational emission in the fundamental band at 4.6~$\mu$m has been detected as close in as 11 au from the star \citep{2009A&A...500.1137V, 2015A&A...574A..75V}. Such a trend of cavity size that increases for larger grain sizes is in quantitative agreement with filtering of the dust by a pressure maximum induced by e.g. an orbiting body \citep[e.g.][]{1979MNRAS.188..191L, 2012A&A...545A..81P} or a dead zone \citep{2015A&A...574A..68F}, although the latter is not expected to leave an imprint in the gas \citep{2016A&A...590A..17R}. There is furthermore a hint of an azimuthal peak in brightness at the south-eastern side of the inner rim of the outer disk show in Fig. \ref{fig:model_comparison}, where the residuals after subtraction of the model show azimuthal asymmetry in the form of a surplus of material at the SE side of the inner disk rim and a deficiency at the other side. Additional higher S/N observations are needed to confirm the presence of this azimuthal brightness asymmetry.

Finally, the strong NIR excess present in the SED is suggestive of hot dust close to the star. \citet{2013A&A...555A..64M} for example model an inner disk between 0.1 and 2 au to explain this emission. The emission from this inner disk is a factor $\approx$ 1000 weaker compared to the contribution of the outer disk at the ALMA band 7 wavelength, and our data cannot confirm nor deny the existence of this inner disk. 

\subsection{Large grains and free-free emission filling in the disk cavity in the ATCA data} \label{disc:free}

The value of the spectral index for the thermal dust emission of 3.1 $\pm$ 0.1 is consistent with a dust distribution in the disk with relatively large grains. These would certainly be up to mm sizes, and thus indicative of significant growth from typical interstellar dust grain sizes of around one tenth of a micron or so {\gc \citep[e.g. ][]{2015ApJ...798...85P}.}

The break in the SED shown in Fig. \ref{fig:SED} around $\approx$ 10~mm is suggestive of emission from another mechanism, additional to the thermal dust emission. Free-free emission (thermal bremsstrahlung) from an idealized spherical, isothermal, constant-velocity outflow is predicted to have a spectral index of $\approx$ 0.6 \citep{1975MNRAS.170...41W}. We repeat that the ALMA band 7 fluxes trace at least partially optically thick emission as suggested by our modeling, and note that the ATCA observations presented in this work cover multiple years. Free-free emission from an ionized wind may vary by $\approx$ 20 to 40 \% on a time-scale of years \citep{2002ApJ...580..459G, 2007ApJ...657..916L}. On the other hand, non-thermal emission may vary over a time-scale of minutes to hours by up to an order of magnitude or more \citep{1986AJ.....92..895K, 1996AJ....111..355C}. Variability has been detected on relatively short time-scales of minutes to hours for e.g. HD~100546 in \citet{2015MNRAS.453..414W}, and for factors of a few as shown e.g. for DK Cha, T Cha and Sz~32 \citep{2012MNRAS.425.3137U}. 

The 5.5/9~GHz observations of 30 May 2011 actually consist of two 30 minute observations separated by about 5 hours. We do not see evidence for variability between the two epochs at either frequency. However, given the considerations previously mentioned concerning the challenges of these frequencies, a definitive statement about variable cm emission for HD~97048 must await additional (and better) data.

\begin{figure}
   \centering
   \includegraphics[width=\hsize]{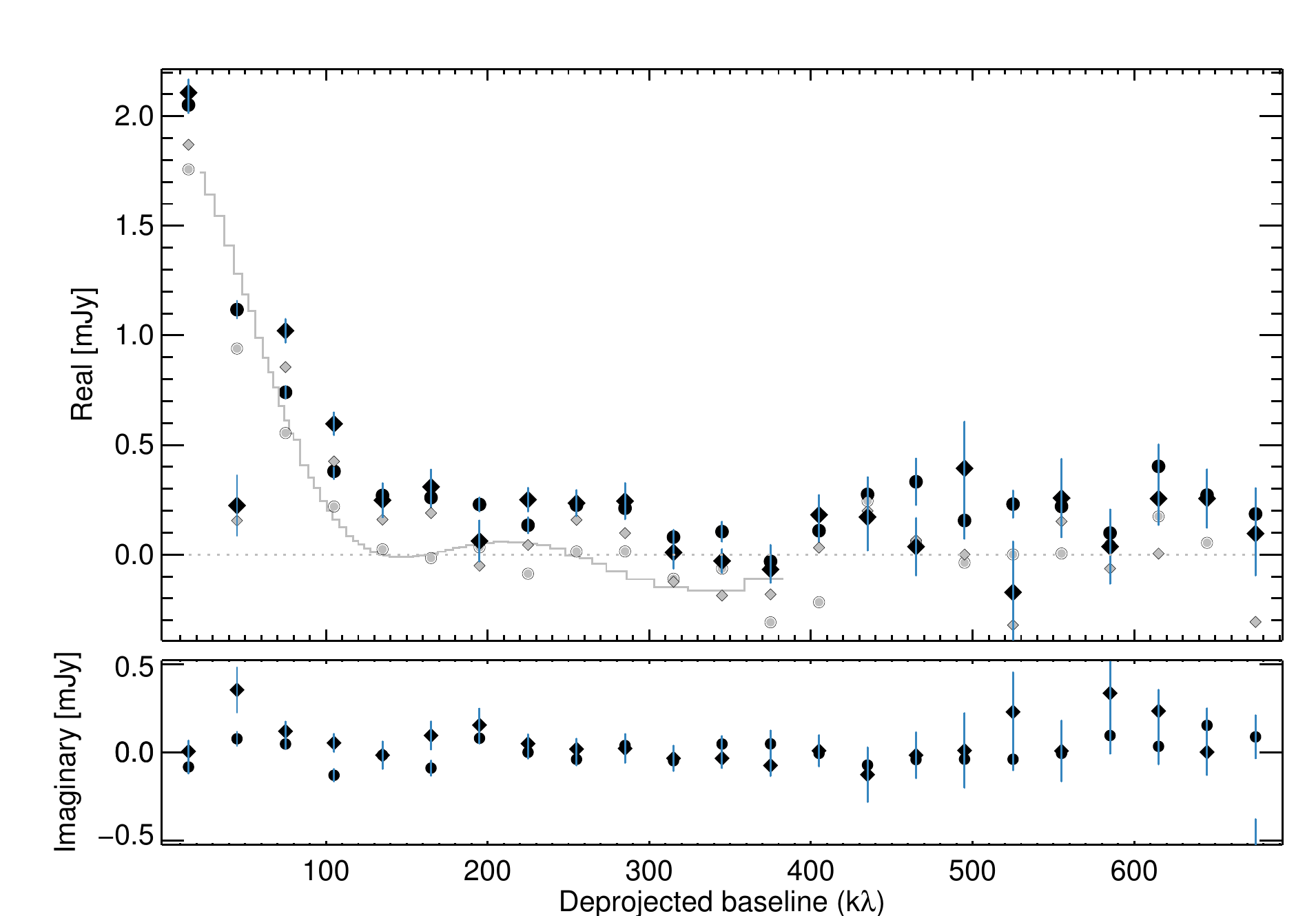}
     \caption{Visibilities of the ATCA 33+35 GHz (circles) and 38+40 GHz (diamonds) observations deprojected using the disk inclination and position angle determined using the ALMA band 7 observations as listed in Table \ref{tab:flux}, and their 1$\sigma$ error values shown in blue. We also show a version of our best-fit MCFOST ring model scaled to the 34 GHZ flux with a grey line and the ATCA visibilities after subtracting a 0.2 mJy point source at the stellar position for the 33+35 GHz observations and a 0.15 mJy point source at the stellar position for the best part of the 38+40 GHz observations, both with grey symbols.}
    \label{fig:ATCA_visibilities}
\end{figure}

Despite their comparable resolutions, the 34~GHz ATCA image does not show the disk cavity as seen in the ALMA band 7 image. Instead, regardless of the weighting used in the inversion and image reconstruction, including natural, uniform and super-uniform, the emission appears centrally peaked at the stellar position within positional uncertainties. Even so, at all of the frequencies observed with ATCA between 33 and 97~GHz, the primary component of the emission is easily resolved. In the most complete data set of 29 August 2014, the source size, inclination and PA are consistent with those properties determined from the ALMA data and what had previously been inferred for the disk from other observations. Yet, as seen in Fig. \ref{fig:ATCA_visibilities}, the real part of the deprojected visibilities does not cross zero. There is no null which would otherwise indicate a relatively sharp drop in the dust density, i.e. a cavity. Instead the visibility drops from the peak of $\geq$ 2~mJy and then plateaus at around 0.15--0.2 mJy. 

Assuming that the main lobe of emission in Fig. \ref{fig:ATCA_visibilities}, and the plateau, represent two different source components, there is about an order of magnitude difference between the two contributors. Since in uv-space the amplitude of a point source has a constant visibility, we suggest that the plateau arises from an unresolved component. This is unlikely to be {\gc either hot dust emission from an inner disk or a photo evaporating inner disk, since the inner disk is} not seen in the ALMA data. So it must instead be either free-free emission from a wind or outflow, or otherwise non-thermal emission. Fitting an unresolved point source to our data in the image plane yields a flux for that component of 0.2 mJy, similar to the plateau component seen in the visibilities of the fourier plane. 

Furthermore, using \textit{uvfit} to simultaneously fit the 33/35~GHz 29 August 2014 data with point and Gaussian sources -- with minimal channel averaging and no binning over uv distance -- yields fluxes of around 0.18 and 1.85~mJy at 33~GHz, and 0.19 and 2.21~mJy at 35~GHz. We note that these nicely reproduce the total fluxes listed in Table \ref{tab:flux}. Similarly for the best data on 30 August 2014, the respective point and Gaussian fluxes are around 0.16 and 2.15~mJy at 38~GHz, and 0.16 and 2.28 mJy at 40~GHz, again consistent with the total fluxes in Table \ref{tab:flux}. Doing the same for the 43/45~GHz observations of 2 August 2009 -- although the uv coverage is sparser and the S/N lower -- suggests the point source component is around 0.3$\pm$0.1~mJy. Models for a `point plus disk' and `point plus ring' were also tried at 33/35 and 38/40~GHz, but the results were unsatisfactory and did not reproduce the observed total fluxes in Table \ref{tab:flux}. To the extent that the inter- and intra-band spectral indices can be believed, it appears that the spectral index of the point source is thus quite flat. 
Interestingly, the relatively low value of the spectral index $\alpha$ beyond 10~mm is in good agreement with the point source flux of the `point plus Gaussian' source model. It may also be consistent with a free-free emission scenario.

Overall, the shape of the ATCA visibilities at both 33/35 and 38/40~GHz are in general agreement with the ALMA band 7 visibilities, inclusive of the fast drop out to 130 k$\lambda$ and the minimum around 350 k$\lambda$. To better compare the ALMA and ATCA data sets, we subtract a point source from the ATCA visibilities using the MIRIAD task {\textit uvmodel}, with point source fluxes as given above. The resulting visibilities, together with a version of the best-fit MCFOST model visibilities scaled to the ATCA 34 GHz flux, are shown in the top panel of Fig. \ref{fig:ATCA_visibilities}. All three visibility curves are in quantitative agreement, which suggests that the disk also has a cavity as traced by grains emitting at 34~GHz. We confirm that adding an unresolved component at the stellar position to our model and convolving it with a beam comparable to that of the ATCA observations fills in the disk cavity in the image plane.

\subsection{Radial drift, a flaring disk structure, gas inside the dust cavity and {\gc apparent} non-Keplerian motions}

The disk around HD~97048 is more spatially extended than the continuum disk in all of the gas lines detected. The continuum disk at 0.853~mm shows a very sharp drop at 1.95\arcsec (Fig. \ref{fig:radial_intensity}), while the \element[ ][12]{CO} J=3-2 emission line is detected out to a radial extent of 5.2\arcsec. Such a sharp drop in the radial continuum brightness and a size discrepancy between a gas line and the associated continuum, are fingerprints for transport of the larger dust grains, dominated by inward radial drift {\gc \cite[e.g.][]{2009A&A...501..269P, 2012ApJ...744..162A}}.

Based on the peak intensity map shown in Fig. \ref{fig:moments-co32} in which the disk opening angle can be seen as a butterfly pattern, it is evident that the \element[ ][12]{CO} J=3-2 emission originating from the disk midplane is weaker than the emission that traces the warm molecular layer higher up in the disk, and is similar to the PAH emission which tracer very small particles in the disk surface,  imaged by \citet{2006Sci...314..621L}. This is more clearly visualized in Fig. \ref{fig:channels}, where we show selected channels maps of the \element[ ][12]{CO} and \element[+][]{HCO} emission. The central panels show projected velocities between $\pm$1.01 and $\pm$1.65 km s$^{-1}$ from the systemic velocity, where the emission is most extended. \element[ ][12]{CO} emission predominantly originating from both the near and far side of the disk can be seen in these panels, separated by the \element[+][]{HCO} emission, which traces emission closer to the midplane. As \element[+][]{HCO} ions quickly disappear without gas-phase CO molecules \citep[e.g.][]{2014ApJ...794..123C}, gas-phase \element[][]{CO} molecules need still be present at the depths where the   \element[+][]{HCO} emission originates. A vertical temperature gradient that cools towards the midplane, as discussed in \citet[e.g.][]{2008ApJ...673L.195S, 2013A&A...557A.133D, 2013ApJ...774...16R} for HD~163296, naturally explains the vertical separation of the \element[ ][12]{CO} emission while still allowing \element[+][]{HCO} emission.

Based both on Keplerian velocities and the location in the resolved maps, the \element[ ][12]{CO} and the \element[+][]{HCO} emissions extend inside the cavity. Calculated from the width of the lines, the \element[+][]{HCO} can be traced down to 27.0 au and the \element[ ][12]{CO} down to 13.4 au. This latter value is in good agreement with the inner radius found for the ro-vibrational CO emission lines that emit from a radius of $\approx$ 11 au \citep{2009A&A...500.1137V, 2015A&A...574A..75V}. Inside of this radius, the disk appears to be devoid of warm CO gas most sensitively traced by ro-vibrational CO emission, but some gas is present, as atomic oxygen gas emission is detected as close in as several tenths of an au \citep{2006A&A...449..267A}. 

Inside the cavity, the velocity map of the \element[+][]{HCO} emission (see Fig. \ref{fig:moments-hcop}) shows an `s' shaped pattern, characteristic for non-keplerian in-falling gas or an inner disk warp, as described by e.g. \citet{2014ApJ...782...62R} and \citet{2015ApJ...811...92C}. This pattern is not visible in the \element[ ][12]{CO} velocity map, which is likely a consequence of the broad foreground absorption. Given the coarse spatial resolution of our observations inside the cavity, we cannot distinguish between both scenarios. {\gc The presence of a small inner disk between 0.3 and 2.0 au as proposed by \cite{2013A&A...555A..64M}, in combination with a} substellar companion in an inclined orbit with respect of the outer disk \citep{2015ApJ...811...92C} could drive a persistent warp of the inner disk. Regardless, either scenario for the {\gc apparent} non-Keplerian motion favours strong gravitational torques inside the cavity to either remove the angular momentum from the gas \citep{2014ApJ...782...62R} or to warp a possible inner disk.

Summarizing, both the cavity size which varies with wavelength and the presence of {\gc apparent} non-Keplerian motions inside the disk cavity, hint at the presence of an orbiting body inside the cavity. This is, together with the very low mass accretion rate, in line with the conclusions of \citet{2015A&A...582L..10K}, who propose that the depletion of heavy elements found in the stellar spectrum emerges as a Jupiter-like planet blocks the accretion of part of the dust and to lesser extent the gas.

\subsection{HD~97048 in context with other transitional disks}

Before ALMA, only the most obvious disk structures, such as inner cavities and very large gaps, could be identified from the SED and/or low-resolution imaging. ALMA provides the combination of resolution and sensitivity needed to reveal more subtle structures, like narrow gaps and rings, azimuthal asymmetries, and spiral arms. Such structures have also been revealed by the new generation of extreme-adaptive optics imagers, including the Gemini Planet Finder and SPHERE \citep{2015ApJ...813L...2W, 2015ApJ...815L..26R}. 

The inner cavity in HD~97048 has previously been inferred from spatially resolved spectroscopic 20 $\mu$m emission and SED modeling \citep{2013A&A...555A..64M}. However, the gap (or ring) in the outer disk was previously unknown.{\gc In addition, \citet{2016arXiv160902011W} infer yet another  peak in the intensity distribution at $\approx$ 300 au, based on modeling of the visibilities.} Similar gaps have been seen in the Class I object HL Tau \citep{2015ApJ...808L...3A} and in the much older TW Hydra disk \citep{2016ApJ...820L..40A}, and might be relatively common considering that only a handful of targets have been observed at comparable resolution. A recent reanalysis of the HL Tau data  suggests the gaps are also present in the gas surface density as traced by the \element[+][]{HCO} J=1-0 line observations \citep{2016ApJ...820L..25Y}. The origin of these gaps is still a matter of intense debate. As discussed in the previous section, proposed explanations include: dynamical clearing by massive planets \citep{2005ApJ...619.1114W}, enhanced particle growth at snow lines \citep{2013ApJ...766...82Z}, and MHD effects \citep{2015A&A...574A..68F}.

{\gc With the caveat that the limited spatial resolution of our observations does not allow us to differentiate between the presence of a gap or a ring, the properties of the disk around HD~97048 }are remarkably consistent with giant planet-formation. Based on the current observational constraints on the occurrence of giant extra-solar planets \citep[e.g.][]{2014ApJ...781...28M, 2015A&A...574A.116R}, it is clear that most protoplanetary disks should not form planets massive enough to open wide gaps. However, demographics of extra-solar planets around main-sequence stars indicate that on average each star has at least 1 planet \citep{2012Natur.481..167C}, and we therefore expect that many disks will form planets. A high incidence of planet-induced gaps in ALMA images is still consistent with the demographics of extra-solar planets given the strong observational bias toward observing the brightest, most massive disks at the resolution required to reveal substructure. Such disks represent the tail of the disk mass distribution and are therefore the most likely birth places of giant planets. We note that even though the TW Hydra disk might currently have a modest luminosity, it is an unusually long-lived disk and it probably was much more massive in the past. Future long-baseline surveys are needed to establish the incidence of narrow gaps in protoplanetary disks and how their occurrence depends on disk properties such as mass and age. 

\subsection{Minimum planetary mass needed to explain the observed disk cavity}

Planetary signatures in disks are more easily detected in the spatial dust distribution than in that of the gas. Dust grains that are coupled to the gas through aerodynamic drag tend to collect in pressure maxima in disks \citep{1995A&A...295L...1B}. If a gap were to be opened by a planet, dust would accumulate in the induced pressure maximum at the outer edge of the planet gap, resulting in  an enhanced contrast between the gap and its outer edge in the dust continuum emission. This contrast in the dust would be greater than that seen in the gas. This enhanced observability has been demonstrating using simulations by e.g. \citet{2006MNRAS.373.1619R, 2014ApJ...789...59O, 2014ApJ...785..122Z}. More recently, \citet{2016MNRAS.459.2790R} estimated the minimum planetary mass needed to produce detectable signatures at different wavelengths in protoplanetary disks, based on the radial location of the planet and the width and distance of the disk gap. We use their simulations to estimate the mass a hypothetical planet should have, were it to be responsible for inducing the cavity we detect at 853 $\mu$m.

Since the sub-mm image shows a cavity, we estimate the hypothesized planet to be located between the outer rim of the inner disk, at 2.5 au \citep{2013A&A...555A..64M}, and the most inner location where CO gas is detected, at 11 au \citep{2009A&A...500.1137V}. These distances yield a ratio between the gap width and gap radius in the range of 0.06 and 0.26. Using a stellar mass of 2.5 M$_\sun$, we estimate a maximum planetary mass between 0.12 and 0.71 M$_{jup}$, based on simulation parameters adopted for Fig.~17 in \citet{2016MNRAS.459.2790R}, and with the caveat that this is a lower limit if the cavity in the HD~97048 disk is younger than 400 planetary orbits.

\subsection{A disk around ISO-ChaI~126} \label{dis:iso-cha1-126}

The spectral index measured between 70 $\mu$m and 15.8 mm of 1.9 $\pm$ 0.2 is consistent with thermal dust emission originating from a disk. The difference between the position angles of the elongated structure and the binary of 55\degr argues against emission from an outflow. We therefore interpret the emission as coming from a disk. Based on the major and minor axis ratio, this disk is inclined by 71\degr$^{+5}_{-7}$. We estimate the emitting dust mass at 43 GHz, by assuming the disk is isothermal and optically thin, using the following relation:
\begin{equation}
\label{eq:mdust}
\centering
\log M_{dust} = \log S_{\nu}+2 \log d-\log \kappa_{\nu}-\log B_{\nu}(\langle T_{dust} \rangle).
\end{equation}
Here, $S_{\nu}$ is the 43 GHz flux density, $d$ is the distance, $\kappa_{\nu}$ is the dust opacity, and $B_{\nu}(\langle T_{dust} \rangle)$ is the blackbody function at the dust temperature. We scale the opacity to 43 GHz with the assumptions of $\kappa_{1.3mm}$=2.3~g$^{-2}$cm$^{2}$ and $\kappa \sim \nu^{0.4}$, following \citet{2013ApJ...771..129A}. We estimate the averaged dust temperature in a disk with an outer radius of 150 au around a 0.5 M$_\sun$, 3~Myr old star, to be 15 K \citep{2016ApJ...819..102V}. The combined mass of the binary system is 1.02$^{+0.58}_{-0.39}$  M$_\sun$ \citep{2013A&A...554A..43D}. Using a gas-to-dust ratio of 100, these assumptions result in a total disk mass of 0.0082 M$_\sun$, or 0.80\% of the total system mass. This is a very reasonable value for the disk and stellar mass ratio, and strengthens our disk interpretation for the resolved structure around ISO ChaI-126.

\section{Conclusions}\label{sec:conclusion}

We summarize the main conclusions of this work as follows:

\begin{itemize} 
        \item The dust disk of HD~97048  is resolved and extends radially out to 2.25\arcsec, after which the surface density drops sharply.
        \item There is a dust cavity visible in the 853 $\mu$m continuum emission out to 43 $\pm$3 au. 
        \item Beyond the outer disk inner rim, we find additional radial structure in the continuum surface brightness profile. This radial structure can be modeled with either a disk gap centered at $\approx$ 90 au \textit{or} an extra emitting ring centered at $\approx$ 150 au. 
        \item The disk cavity is not detected at 9~mm because free-free emission from the star fills in the cavity at our spatial resolution. 
        \item The \element[ ][12]{CO} J=1-0, \element[ ][12]{CO} J=3-2 and \element[+][]{HCO} 4-3 emission lines all are more extended than the disk continuum emission, up to a radius of 5.2\arcsec for the \element[ ][12]{CO} J=3-2 emission. The discrepancy in size between the mm dust emission and the associated line emission, together with the sharp outer edge of the dust disk, can be explained by inward radial drift of the larger dust grains. 
        \item Both the \element[+][]{HCO} J=4-3 and the \element[ ][12]{CO} gas extend inside the dust cavity, down to distances as close as 13.4 au from the central star.
        \item The \element[+][]{HCO} intensity-weighted velocity map shows {\gc a velocity structure deviating from Keplerian motion expected from a co-planar disk} inside of the cavity. {\gc Possible explanations for this velocity structure are a distortion by an inclined inner disk and an extra, non-Keplerian, velocity component.}
        \item A cavity size varying with dust grain size and the {\gc apparent} non-Keplerian motions inside of the cavity both hint at an extra body orbiting inside the cavity. 
        \item A planet with a mass of $\approx$ 0.7 M$_{jup}$ at a radial distance between 2.5 and 11 au could open up the disk cavity observed at 853 $\mu$m. 
        \item We resolve emission between 2.9 and 16.7 mm from the 0.5~M$_{\odot}+0.5$~M$_{\odot}$ binary ISO-ChaI 126 that we interpret as originating from a disk. The spectral slope of this emission between 70 $\mu$m and 15.8 mm is 1.9 $\pm$ 0.2. Using standard assumptions, we calculate a total disk mass of 0.80\% of the total system mass.
    \end{itemize}

\begin{acknowledgements}

GvdP, SC, LC, SP and HA acknowledge support from the Millennium Science Initiative (Chilean Ministry of Economy) through grant RC130007. GvdP acknowledges financial support from FONDECYT, grant 3140393, CMW acknowledges financial support from an Australian Research Council Future Fellowship FT100100495, LC acknowledges financial support from FONDECYT, grant 1140109, S.P. acknowledges financial support by FONDECYT, grant 3140601, HA acknowledges financial support by FONDECYT, grant 3150643, and SC acknowledges support from FONDECYT grant 1130949. H.C. acknowledges support from the Spanish Ministerio de Econom\'ia y Competitividad under grant AYA2014-55840P. This  paper  makes  use  of  the following ALMA data: ADS/JAO.ALMA\# 2013.1.00658.S and 2012.1.00031.S. ALMA is a partnership of ESO (representing its member states), NSF (USA) and NINS (Japan), together with NRC (Canada) and NSC and ASIAA (Taiwan), in cooperation with the Republic of Chile. The Joint ALMA Observatory is operated by ESO, AUI/NRAO and NAOJ. The National Radio Astronomy Observatory is a facility of the National Science Foundation operated under cooperative agreement by Associated Universities, Inc. 
\end{acknowledgements}

\begin{appendix}
\section{Complete channel maps for the  \element[ ][12]{CO} J=3-2 and \element[+][]{HCO} J=4-3 emission lines}

We show a full set of the channel maps for the  \element[ ][12]{CO} J=3-2 (Figs. \ref{fig:channels-co-1} - \ref{fig:channels-co-3}) and \element[+][]{HCO} J=4-3 (Figs. \ref{fig:channels-hco-1} and \ref{fig:channels-hco-2}) emission lines in this appendix. The spectral resolution is 106 m s$^{-1}$ per channel for the  \element[ ][12]{CO} J=3-2 emission and 103 m s$^{-1}$ per channel for the  \element[+][]{HCO} J=4-3 emission.

\begin{figure*}
   \centering
   \includegraphics[width=0.87\textwidth]{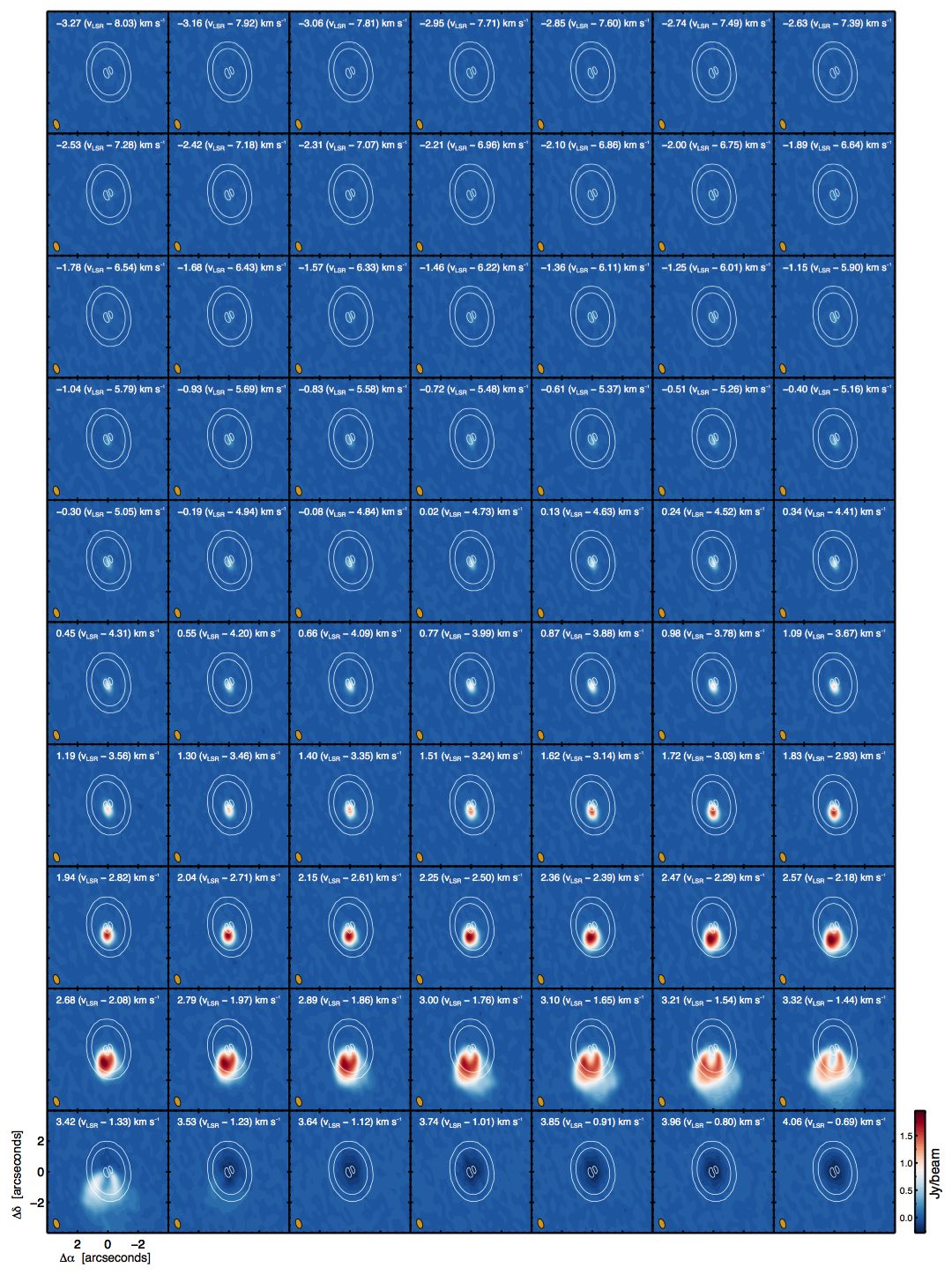}
     \caption{Channel maps showing the \element[ ][12]{CO} J = 3-2 spectral channels at 106 m s$^{-1}$ resolution between +8 and -8 km s$^{-1}$ around the systemic velocity. In the top right of each panel we note the v$_{lsr}$ in white with the velocity with respect to the systemic velocity of 4.75 km s$^{-1}$ in parenthesis. The clean beam is shown in orange in the bottom left of each panel.  The continuum contours in white.} 
    \label{fig:channels-co-1}
\end{figure*}

\begin{figure*}
   \centering
   \includegraphics[width=0.87\textwidth]{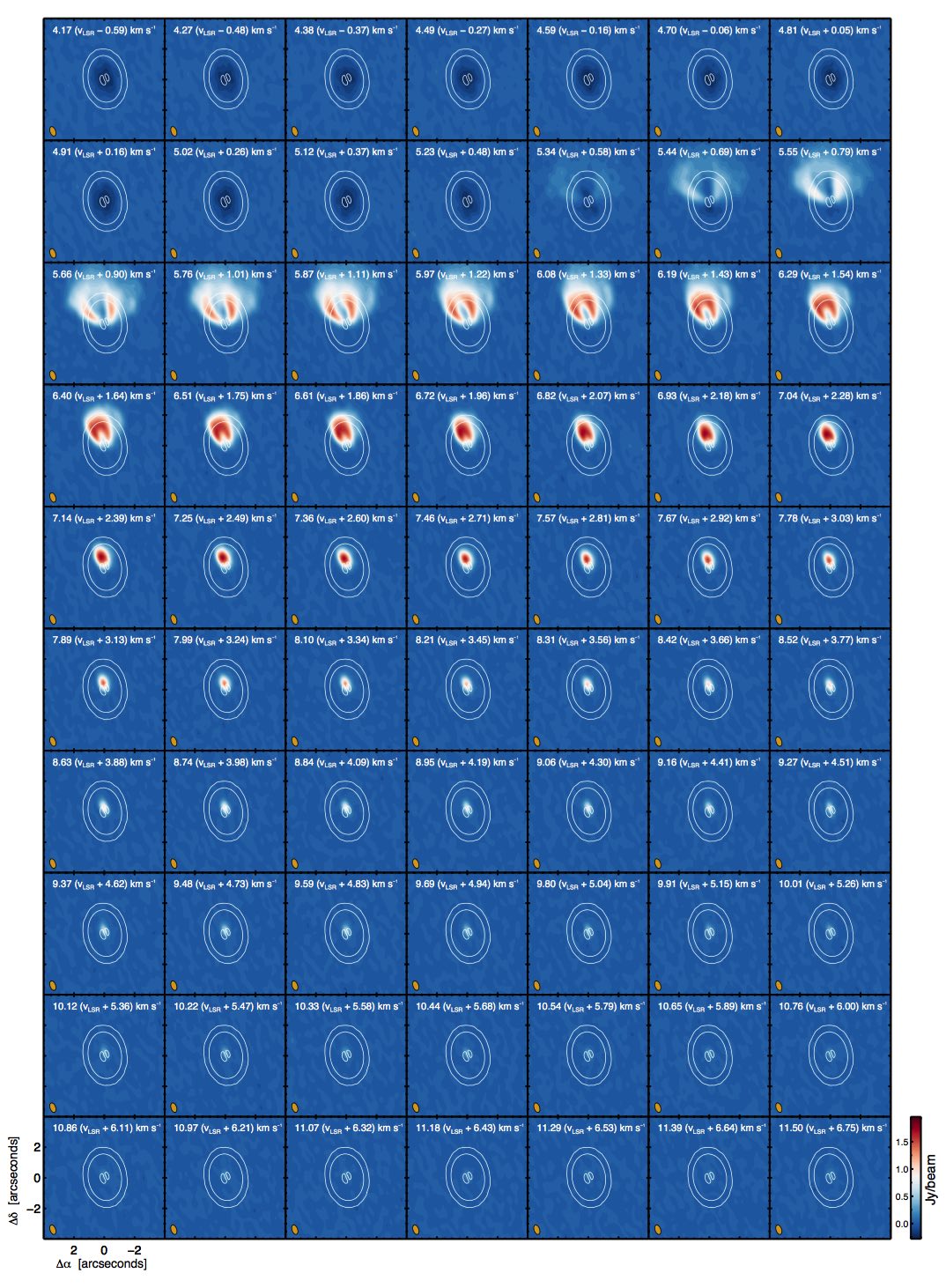}
     \caption{Continuation of Fig. \ref{fig:channels-co-1}.} 
    \label{fig:channels-co-2}
\end{figure*}

\begin{figure*}
   \centering
   \includegraphics[width=0.87\textwidth]{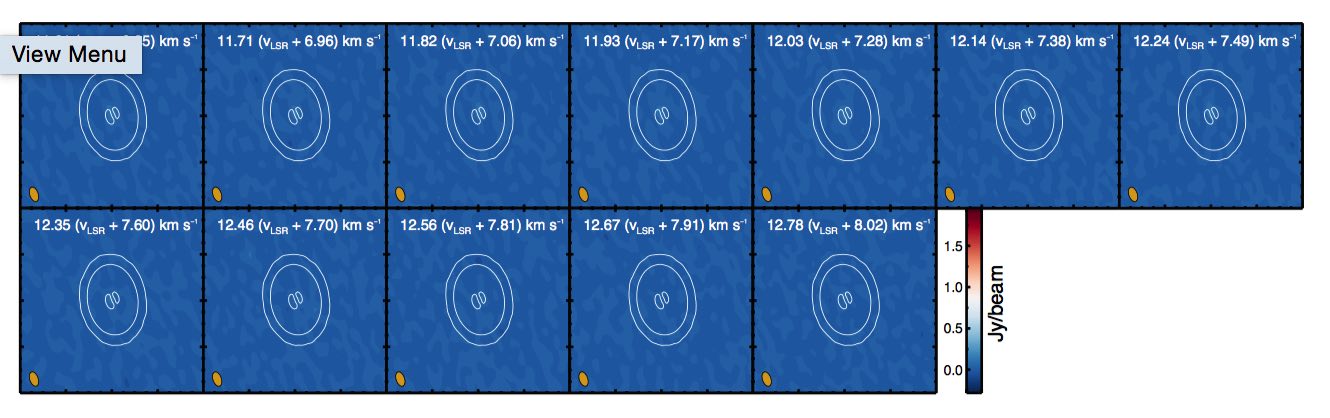}
     \caption{Continuation of Figs. \ref{fig:channels-co-1} and \ref{fig:channels-co-2}.} 
    \label{fig:channels-co-3}
\end{figure*}

\begin{figure*}
   \centering
   \includegraphics[width=0.87\textwidth]{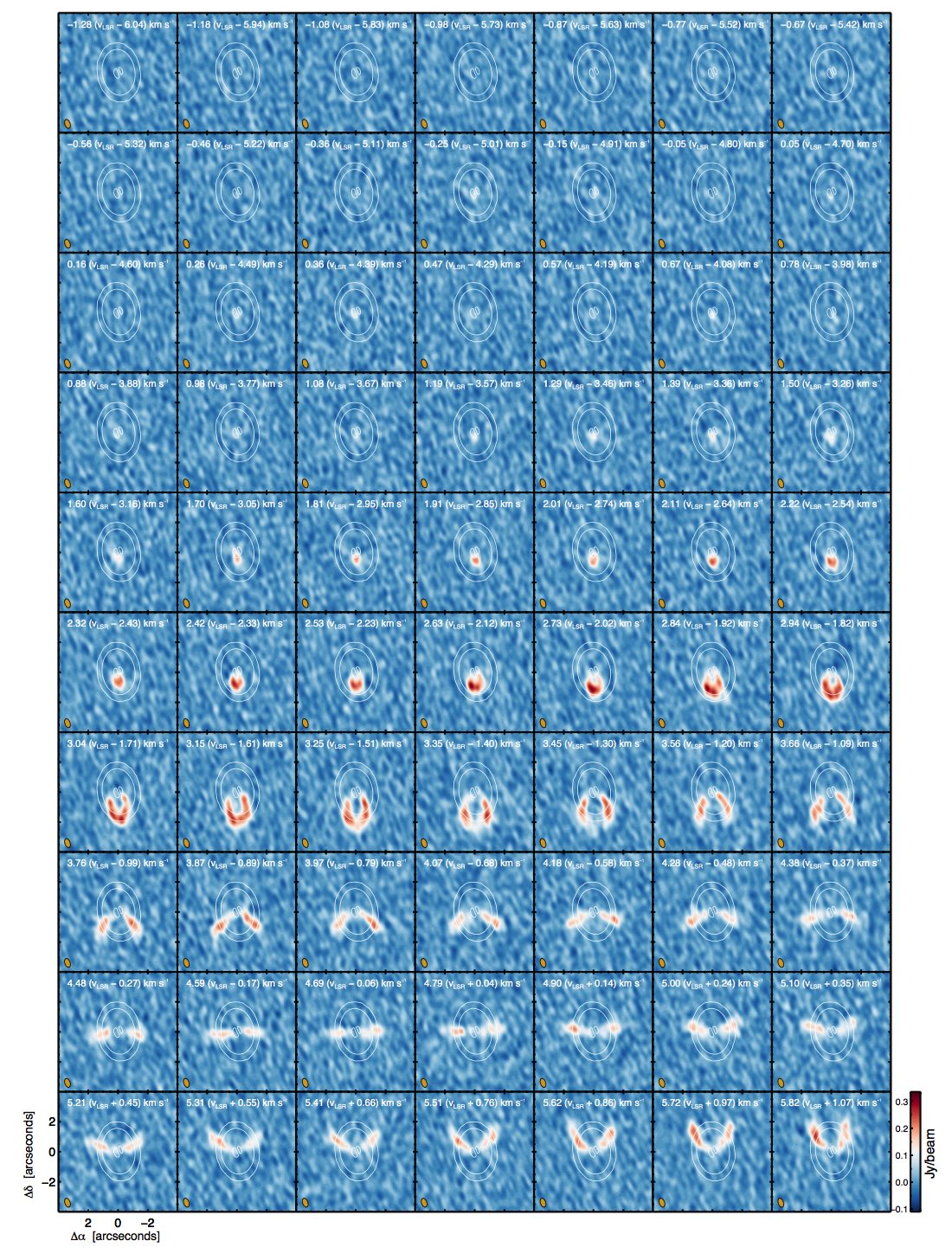}
     \caption{Channel maps showing the\element[+][]{HCO} J=4-3 spectral channels at 103 m s$^{-1}$ resolution between +6 and -6 km s$^{-1}$ around the systemic velocity. In the top right of each panel we note the v$_{lsr}$ in white with the velocity with respect to the systemic velocity of 4.75 km s$^{-1}$ in parenthesis. The clean beam is shown in orange in the bottom left of each panel.  The continuum contours in white.} 
    \label{fig:channels-hco-1}
\end{figure*}

\begin{figure*}
   \centering
   \includegraphics[width=0.87\textwidth]{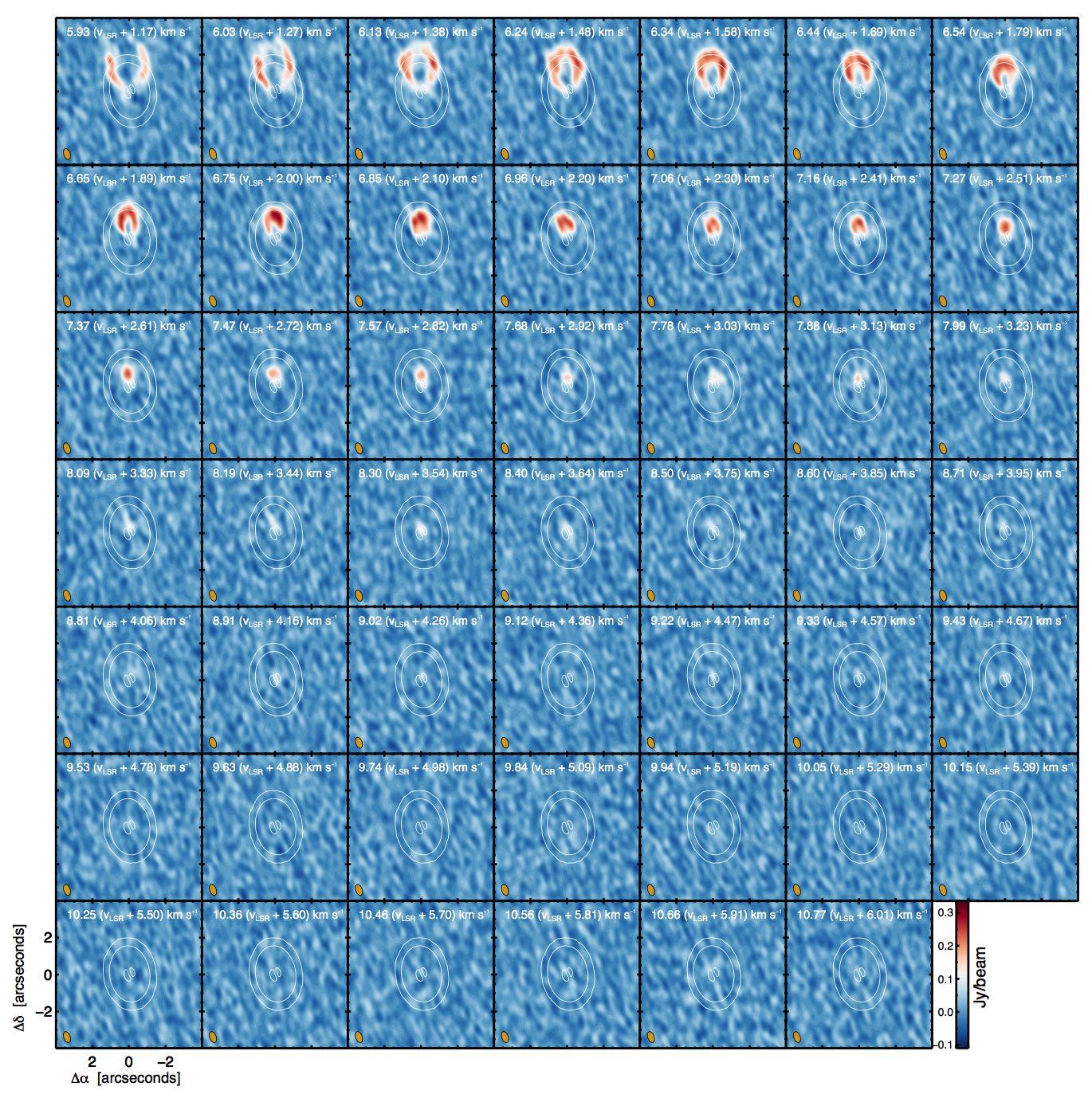}
     \caption{Continuation of Fig. \ref{fig:channels-hco-1}.} 
    \label{fig:channels-hco-2}
\end{figure*}

\end{appendix}

\end{document}